\documentclass[preprint,A4]{elsarticle}

\usepackage[width=15cm,height=22cm,left=2.75cm, top=2.75cm]{geometry}
\usepackage{placeins}  
\usepackage{float}     
\usepackage{verbatim}  
\usepackage{xspace}    
\usepackage{changepage} 

\usepackage{subcaption}
\usepackage{graphicx} 
\usepackage{graphics} 
\usepackage{booktabs} 

\usepackage{lineno,hyperref}
\modulolinenumbers[5]
\usepackage{glossaries} 
\makeglossaries

\usepackage{amsmath} 

\usepackage{lipsum} 

\journal{Journal of \LaTeX\ Templates}

\newacronym{CR}{CR}{Cosmic Ray}
\newacronym{CRs}{CRs}{Cosmic Rays}
\newacronym{UHECR}{UHECR}{Ultra High Energy Cosmic Rays}
\newacronym{UHE}{UHE}{Ultra High Energy}
\newacronym{SEP}{SEP}{solar energetic particles}
\newacronym{GCR}{GCR}{Galactic Cosmic Rays}
\newacronym{ACR}{ACR}{Anomalous cosmic rays}
\newacronym{GZK}{GZK}{Greisen-Zatsepin-Kuzmin}
\newacronym{CMB}{CMB}{Cosmic Microwave Background Radiation}
\newacronym{UV}{UV}{Ultra-Violet radiation}
\newacronym{IR}{IR}{Infra-Red radiation}
\newacronym{Opt}{Opt}{Optical radiation}
\newacronym{URB}{URB}{Cosmic Universal Radio Radiation} 
\newacronym{SNR}{SNRs}{Supernova Remnants}
\newacronym{AGN}{AGN}{Active Galactic Nuclei}
\newacronym{VCV}{VCV}{V\'{e}ron-Cetty and V\'{e}ron catalogue}
\newacronym{ICS}{ICS}{Inverse Compton Scattering}

\newacronym{RFT}{RFT}{Gribov's Reggeon Field Theory}
\newacronym{LPM}{LPM}{Landau-Pomeranchuk-Migdal effect}
\newacronym{QCD}{QCD}{Quantum Chromodynamics}
\newacronym{DPM}{DPM}{Dual Parton Model}
\newacronym{EPOS}{E{\textsc{pos}}}{Energy conserving quantum mechanical multiple scattering approach, based on Partons, Off-shell remnants and Splitting of partons ladders}
\newacronym{QGSjet}{QGSJ{\textsc{et}}}{Quark Gluon String Model with mini-Jet}
\newacronym{Fluka}{FLUKA}{Gamma-Hadron-Electron-Interaction SH(A)ower code}
\newacronym{Gheisha}{Gheisha}{FLUktuierende KAskade (German for 
fluctuating cascade)}
\newacronym{CORSIKA}{CORSIKA}{COsmic Ray SImulations for KAscade}
\newacronym{AIRES}{AIRES}{AIRshower Extended Simulations}

\newacronym{CERN}{CERN}{European Organization for Nuclear Research (a particle physics laboratory in Geneva, Switzerland)}
\newacronym{LHC}{LHC}{Large Hadron Collider}
\newacronym{RHIC}{RHIC}{Relativistic Heavy Ion Collider (at BNL)}
\newacronym{TOTEM}{TOTEM}{TOTal Elastic and diffractive cross section Measurement}

\newacronym{KASCADE}{KASCADE}{KArlsruhe Shower Core and Array DEtector}
\newacronym{HiRes}{HiRes}{High Resolution Fly's Eye detector}
\newacronym{TA}{TA}{Telescope Array}
\newacronym{AGASA}{AGASA}{Akeno Giant Air Shower Array}

\newacronym{PAO}{PAO}{Pierre Auger Observatory}
\newacronym{HEAT}{HEAT}{High Elevation Auger Telescope}
\newacronym{AMIGA}{AMIGA}{Auger Muons and Infill for the Ground Array}
\newacronym{AERA}{AERA}{Auger Engineering Radio Array}
\newacronym{MIDAS}{MIDAS}{MIcrowave Detection of Air Showers}
\newacronym{AMBER}{AMBER}{Air-shower Microwave Bremsstrahlung Experimental Radiometer}
\newacronym{EASIER}{EASIER}{Extensive Air Shower Identification using Electron Radiometers}
\newacronym{LOPES}{LOPES}{LOFAR Prototype Station}
\newacronym{LOFAR}{LOFAR}{Low Frequency ARray}
\newacronym{CROME}{CROME}{Cosmic Ray Observation via Microwave Emission}
\newacronym{CODALEMA}{CODALEMA}{COsmic ray Detection Array with Logarithmic ElectroMagnetic Antennas}
\newacronym{AIRFLY}{AIRFLY}{AIR FLuorescence Yield}

\newacronym{rpc}{RPC}{Resistive Plate Chambers}
\newacronym{marta}{MARTA}{Muon Auger RPC for the Tank Array}

\newacronym{CTA}{CTA}{Cherenkov Telescope Array}
\newacronym{MAGIC}{MAGIC}{Major Atmospheric Gamma-ray Imaging Cherenkov}
\newacronym{HESS}{HESS}{High Energy Stereoscopic System}
\newacronym{VERITAS}{VERITAS}{Very Energetic Radiation Imaging Telescope Array System}
\newacronym{HAWC}{HAWC}{the High-Altitude Water Cherenkov Observatory}
\newacronym{SUGAR}{SUGAR}{The Sydney University array}
\newacronym{MACFLY}{MACFLY}{Measurement of Air Cherenkov and Fluorescence Light Yield}

\newacronym{EAS}{EAS}{Extensive Air Shower}
\newacronym{USP}{USP}{Universal Shower Profile}
\newacronym{ldf}{LDF}{Lateral Distribution Function}
\newacronym{MC}{MC}{Monte Carlo}
\newacronym{MPD}{MPD}{Muon Production Depth}
\newacronym{VEM}{VEM}{Vertical Equivalent Muon}
\newacronym{em}{EM}{Electromagnetic signal component}
\newacronym{mu}{MU}{Muonic signal component}
\newacronym{tot}{TOT}{Total signal component}
\newacronym{CIC}{CIC}{Constant Intensity Cut}
\newacronym{NKG}{NKG}{Nishimura, Kamata and Greisen equation}
\newacronym{FLY}{FLY}{Fluorescence Light Yield}

\newacronym{wct}{WCT}{Water-Cherenkov tanks} 
\newacronym{wcd}{WCD}{Water Cherenkov detector} 
\newacronym{PMT}{PMT}{Photomultiplier Tube}
\newacronym{sd}{SD}{Surface Detectors}
\newacronym{fd}{FD}{Fluorescence Detectors}
\newacronym{fov}{FOV}{Field Of View}
\newacronym{SDP}{SDP}{Shower Detector Plane}
\newacronym{FADC}{FADC}{Flash Analog to Digital Converter}
\newacronym{ADC}{ADC}{Analog Digital Converter}
\newacronym{GPS}{GPS}{Global Positioning System}
\newacronym{DAQ}{DAQ}{Data Acquisition}
\newacronym{XML}{XML}{eXtensible Markup Language}

\newacronym{CLF}{CLF}{Central Laser Facility}
\newacronym{XLF}{XLF}{Extreme Laser Facility}
\newacronym{cdas}{CDAS}{Central Data Acquisition System}
\newacronym{LIDAR}{LIDAR}{Light Detection And Ranging}
\newacronym{ham}{HAM}{Horizontal Attenuation Monitors}
\newacronym{FRAM}{FRAM}{tttttt}
\newacronym{fram0}{FRAM}{The Photometric Robotic Atmospheric Monitor}
\newacronym{apf}{APF}{Aerosol Phase Function}
\newacronym{FPGA}{FPGA}{Field-Programmable Gate Array}

\newacronym{UrQMD}{UrQMD}{Ultra-relativistic Quantum Molecular Dynamics model}
\newacronym{QGS}{QGS}{Quark-Gluon-String}
\newacronym{CL}{CL}{Confidence Level}

\newacronym{VHF}{VHF}{Very High Frequency radiation}
\newacronym{MBR}{MBR}{molecular bremsstrahlung radiation}
\newacronym{SOPHIA}{SOPHIA}{Simulations Of Photo Hadronic Interactions in Astrophysics}

\newacronym{PLD}{PLD}{Programmable Logic Device}
\newacronym{FE}{FE}{front-end}
\newacronym{CCD}{CCD}{charge-coupled device} 

\newacronym{FLT}{FLT}{First Level Trigger in FD}
\newacronym{SLT}{SLT}{Second Level Trigger in FD}
\newacronym{TLT}{TLT}{Third Level Trigger in FD}
\newacronym{T1}{T1}{First Level Trigger in SD} 
\newacronym{T2}{T2}{Second Level Trigger in SD} 
\newacronym{T3}{T3}{Third Level Trigger in SD} 
\newacronym{T4}{T4}{Fourth Level Trigger (of physics trigger) in SD} 
\newacronym{T5}{T5}{Fifth Level Trigger (or fiducial trigger) in SD} 
\newacronym{TH}{TH}{T1 Simple Threshold trigger in SD} 
\newacronym{ToT}{ToT}{Time-over-Threshold trigger in SD} 

\newcommand{\epos}{EPOS-LHC\xspace}
\newcommand{\qgs}{QGSJ{\textsc{et}}-II.04\xspace}

\newcommand{\DDz}{_{\delta=0} }
\newcommand{\DDzMU}[1]{_{#1,\delta=0} }
\newcommand{\DD}[1]{_{\delta #1}}
\newcommand{\DDsm}[2]{_{#1,\delta #2}}
\newcommand{\DDsmZ}[1]{_{#1}}

\begin{document}

\begin{frontmatter}
\title{Sensitivity of EAS measurements to the energy spectrum of muons}

\author[LIPadress]{J. Espadanal \corref{mycorrespondingauthor}}
\cortext[mycorrespondingauthor]{Corresponding author}
\ead{jespada@lip.pt}
\author[LIPadress]{L. Cazon}
\ead{cazon@lip.pt}

\author[LIPadress,Santiago]{R. Concei\c{c}\~{a}o}
\ead{ruben@lip.pt}

\address[LIPadress]{LIP, Av. Elias Garcia, 14-1, 1000-149 Lisboa, Portugal}
\address[Santiago]{{Depto. de F\'{i}sica de Part\'{i}culas \& Instituto Galego de F\'{i}sica de Altas Enerx\'{i}as, Universidade de Santiago de Compostela, 15782 Santiago de Compostela, Spain}}

\begin{abstract}


We have studied how the energy spectrum of muons at production affects some of the most common measurements related to muons in extensive air shower studies, namely, the number of muons at the ground, the slope of the lateral distribution of muons, the apparent muon production depth, and the arrival time delay of muons at ground.
We found that by changing the energy spectrum by an amount consistent with the difference between current models (namely \epos and \qgs), the muon surface density at ground increases  $5\%$ at $20^\circ$ zenith angle and $17\%$ at $60^\circ$ zenith angle.  This effect introduces a zenith angle dependence on the reconstructed number of muons which might be experimentally observed. The maximum of the muon production depth distribution at $40^\circ$ increases $\sim10\text{ g/cm}^2$ and $\sim0\text{ g/cm}^2$ at $60^\circ$, which, from pure geometrical considerations, increases the arrival time delay of muons. There is an extra contribution to the delay due to the subluminal velocities of muons of the order of $\sim3$ ns at all zenith angles. Finally, changes introduced in the logarithmic slope of the lateral density function are less than 2\%.


\end{abstract}


\end{frontmatter}

\section{Introduction}

In recent years, a new generation of experiments has measured with unprecedented precision the energy spectrum of \gls{UHECR}, demonstrating the existence of a cutoff at the GZK energies \cite{AugerSpectrum,Hires}. However, it is not yet possible to determine the mass of those particles with enough precision to constrain the astrophysical scenarios for their origin, and thus distinguish the origin of such a cutoff, which can be due to the exhaustion of the energy at the sources or to the interaction of the \gls{UHECR} with the \gls{CMB}.

\gls{UHECR} interact with atmospheric atomic nuclei producing a cascading process of particle reactions commonly designated by Extensive Air Shower (EAS). EAS can be measured by detecting the radiation emitted by their passage through the atmosphere or by sampling the secondary particles at ground. 
The reconstruction of the primary \gls{UHECR} properties relies on our understanding of EAS physics and, in particular, of hadronic particle interactions. Our knowledge of hadronic interactions is limited, which is translated into 
phenomenological models including parameters supported by experiments only up to LHC energies.
Moreover,  the available accelerator data do not cover the full kinematic region of interest, namely the forward region, nor the full possible interaction systems such as pion-Nitrogen
interactions, which are of the utmost importance for the description of the shower development.
In summary, the different compositions of the UHECR, uncertainties in the hadronic interactions extrapolations, and the possibility of new physics scenarios at the highest energies often share the same phase-space of EAS observables, making it difficult to disentangle hadronic physics effects from UHECR mass determination. 
The Pierre Auger Collaboration\cite{PAO_0} interprets the evolution of the electromagnetic shower maximum with increasing energy as an indication towards a heavier primary composition \cite{Xmax2014,XmaxInt2014}.
On the other hand, hadronic interaction models fail to accurately predict the number of muons that reach the ground. At $10^{19}$ eV the mean muon number in simulations would have to be increased by $30\%$ (\epos) to $80\%$ (\qgs) \cite{MuonN2014} to match the number of muons deduced from the depth of the electromagnetic shower maximum. Moreover, measurements of the depth at which the muon production rate reaches its maximum at energies above $10^{19.2}$ \cite{MPD2014} are strongly inconsistent with \epos predictions.

In each hadronic interaction, the produced $\pi^0$ carry around $25\%$ of the energy. They decay almost instantly into photons feeding the electromagnetic cascade. The rest of the energy is carried mainly by charged pions and also other mesons and barions, which continue the hadronic cascade. They eventually decay into muons whenever their energy drops below their critical energy, $ \mathcal{O}$(100 GeV).  In this way, the muons act as tracers of the hadronic development: their study gives access to the details of the hadronic cascade.  Assuming that the composition can be properly understood, a discrepancy between observed and predicted muon numbers, might indicate a change in the fraction of energy carried by the neutral pions, hadronic cross sections, or hadronic multi-particle production.

 This paper explores a complementary scenario: a change in the energy distribution of the muons at production (leaving the total number of produced muons unaltered) might produce changes in the muon measurements performed at ground. In particular, it could mimic (or contribute to changes in) the number of muons arriving at ground. Note that the energy distribution of the muons is directly linked to the angular distribution of muon trajectories, the probability of decay, the interactions with the geomagnetic field and Coulomb scattering. In summary, we study the sensitivity of muon observables at ground to changes in the energy of muons at production, keeping their overall number constant.

The paper is organised as follows. In section \ref{section:Procedure}, we describe the muon distributions at production. In section \ref{subsection: Muons Energy}, we present the details regarding the modification of the muon energy spectrum and its propagation to the ground. The results obtained for the muon observables at ground are shown and discussed in sections \ref{section: Number of muons} and \ref{section: Signal of muons}. Finally, we summarise all results in section \ref{section: Sensitivity}.

\section{Distribution of muons at production}
\label{section:Procedure}

According to \cite{CazonTransportModel,Cazon2004}, the positions where muons are produced (where the parent mesons decay) are confined to a relatively narrow cylinder, of a few tens of meters, around the shower axis.
Thus, muons can be approximated as being produced along the shower axis every d$X$ (depth in g/cm$^{2}$ of crossed atmosphere), within a given energy and transverse momentum interval $\text{d}E_i$ and $\text{d}p_t$, respectively:
\begin{equation}
\frac{\text{d}^3N}{\text{d}X \text{d}E_i \text{d}cp_t} = F(X, E_i, cp_t) \:.
\label{eq: dNdXdE}
\end{equation}
In \cite{CazonTransportModel}, it was proven that $F(X, E_i, cp_t)$ contains all the relevant information needed to fully determine the distributions at the ground. Using the transport model described also in \cite{CazonTransportModel}, it was thus possible to fast simulate the effects on the observed muon distributions at the ground after introducing modifications at production through modifying  $F(X, E_i, cp_t)$.

The total number of muons produced in a shower is given by
\begin{equation}
N_0 = \int F(X, E_i, cp_t) \text{d}E_i \text{d}cp_t \text{d}X \:.
\label{eq: hX}
\end{equation}
The projection $h(X)= \int F(X, E_i, cp_t) \text{d}E_i \text{d}cp_t$ is the so called \textit{total/true} \gls{MPD} distribution. 
It represents the production rate of muons in $\mathrm{g/cm^{2}}$. Its shape and features are discussed in \cite{MuonUSP}. This distribution gives the position where the charged pions decay, which means that it tracks the longitudinal development of the hadronic cascade. The \textit{apparent} MPD-distribution, the one that is experimentally accessible, corresponds to the distribution of the production depths of muons that arrived at a given location on the ground after propagation, and thus would be affected by changes in the energy spectrum. The depth at which the apparent MPD distribution reaches the maximum is denoted as $X^{\mu}_{max}$. Note that, only the apparent $X^{\mu}_{max}$ is experimentally accessible, so in this paper, it will correspond to the apparent maximum. 


In this paper, a library of showers simulated with CORSIKA (v7.4100) \cite{CORSIKA} was created, and each individual  $F(X, E_i, cp_t)$ distribution recorded. The library contains samples of 100 showers for each primary (proton and iron) at zenith angles ($\theta$) $0^\circ$, $20^\circ$, $40^\circ$, $60^\circ$ and $70^\circ$, and energy $10^{19}$ eV, with the models \epos and \qgs.
The kinetic energy cuts were set to 0.05 GeV for muons and hadrons, 
and the altitude was set to 1400 m a.s.l., with the magnetic field of Malarg\"{u}e, Argentina, site of the Pierre Auger Observatory.   

\FloatBarrier
\begin{figure}[h]
\centering
\hspace{-0.5cm}
\begin{adjustwidth}{-0.50cm}{-0.50cm}
       \begin{subfigure}[b]{0.5\linewidth}\centering
\includegraphics[width=1\textwidth]{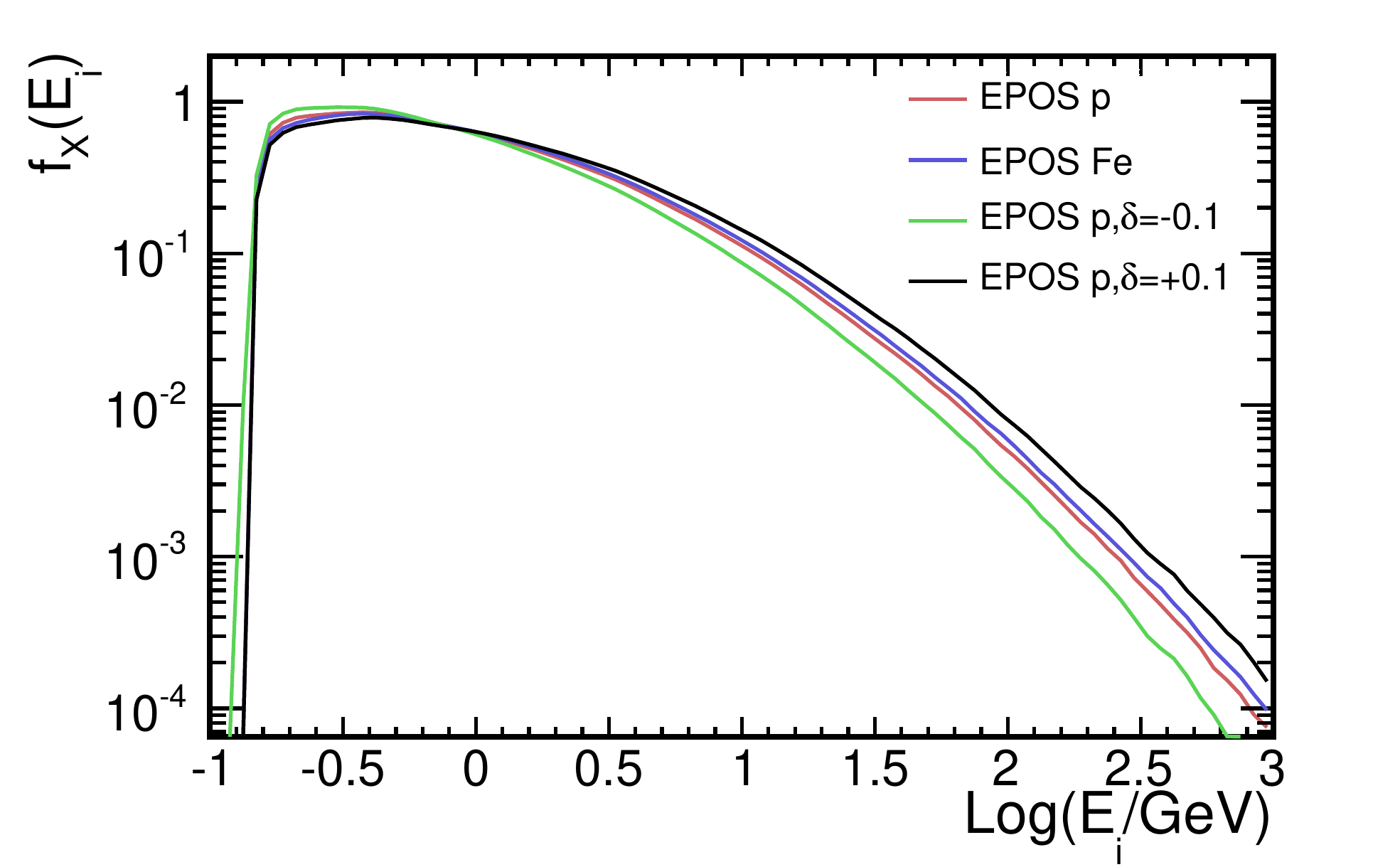}
                \caption{ Muon Spectrum $f_{X}(E_i)$ }
                \label{fig: MuSpectrum a}
        \end{subfigure}%
       \begin{subfigure}[b]{0.5\linewidth}\centering
\includegraphics[width=1\textwidth]{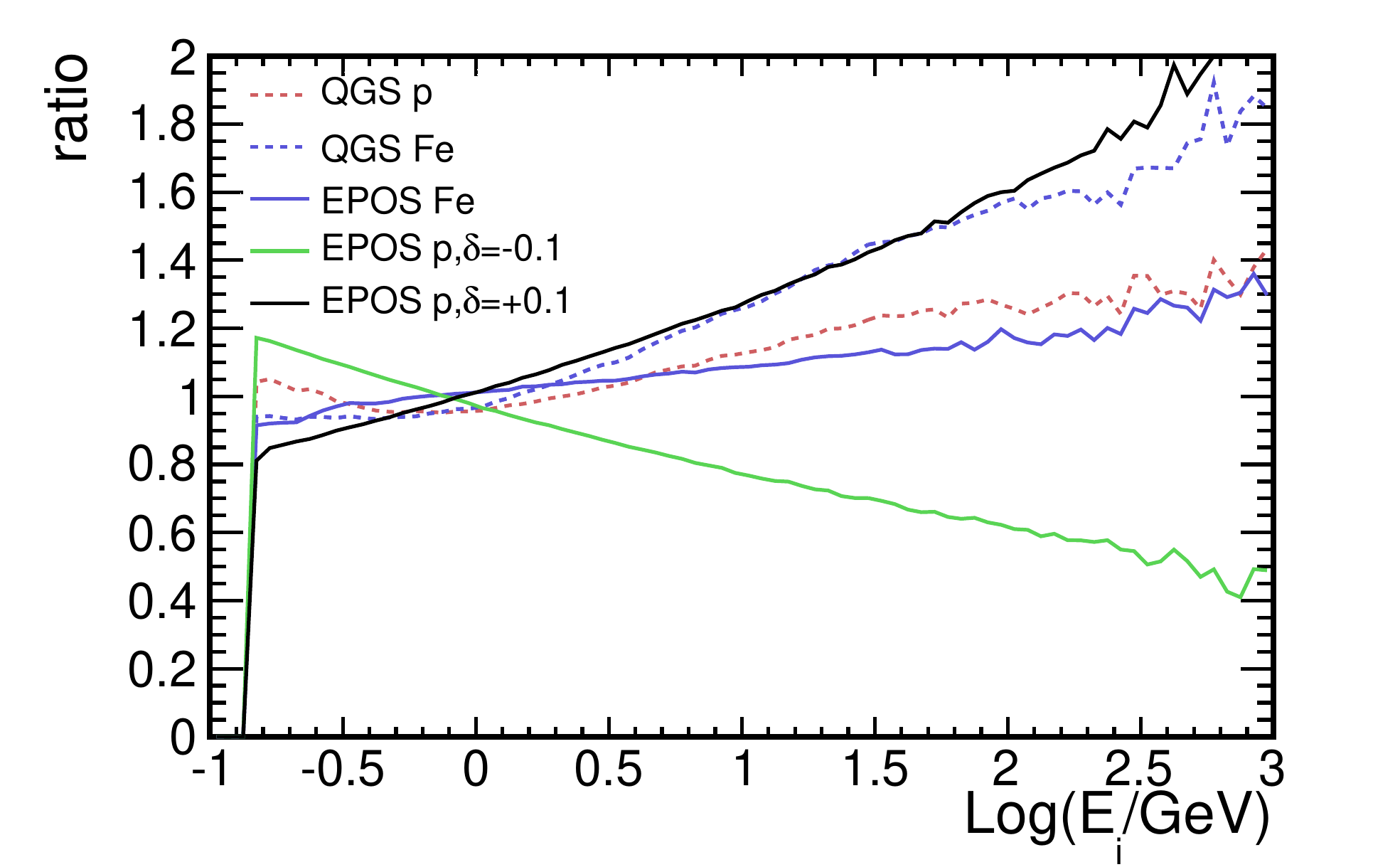}
                \caption{ $f_{X}(E_i)/f_{X,p, EPOS}(E_i)$}
                \label{fig: MuSpectrum b}
        \end{subfigure}%
\end{adjustwidth}
\caption[]{Muon production spectrum at the depth $X\in \left[X^\mu_{max} -25,X^\mu_{max}+25\right] \mathrm{ g/cm^{2}}$, in $40^\circ$ showers (a) and fraction of the spectrum with respect to the \epos proton model (b). Results shown for proton (red) and iron (blue), \epos (dashed) and \qgs (full) models and to \epos proton model considering $\delta=-0.1$ (dashed gray) and $\delta=+0.1$ (dashed green), from eq. \ref{eq: dist}. 
\label{fig:MuSpectrum}}
\end{figure}

\section{Modification of the muon energy spectrum}
\label{subsection: Muons Energy}

The shape of the transverse momentum distribution of muons within EAS is universal, and  does not depend on the primary particle type, zenith angle, or hadronic interaction model \cite{CazonTransportModel}.  Therefore we have chosen 
to study the effects of changes in the energy of muons at the production point by scaling $| \vec{p}|$ or $E_i$ while maintaining the transverse momentum $p_t$ constant. This necessarily implies a change in the angle of the 3-momentum with respect to the shower axis $\alpha$, given that $\sin \alpha=\frac{cp_t}{E_i}$.

For a given depth $X$, the energy distribution $E_i$ can be given by:
\begin{equation}
f_{X}(E_i) = \frac{\text{d}^2N}{\text{d}E_i\text{d}X} =\int F(X,E_i,cp_t) \text{d} cp_t \: .
\end{equation}
In figure \ref{fig: MuSpectrum a}, $f_{X}(E_i)$ is plotted for the depth $X\in \left[X^\mu_{max} -25,X^\mu_{max}+25\right] \mathrm{ g/cm^{2}}$, while in figure \ref{fig: MuSpectrum b}, the muon spectrum of proton and iron \qgs and iron \epos is compared to the one from proton \epos. The spectrum is harder for iron primaries than for proton primaries and is harder in the case of \qgs. 

 The high energy tails of the muon spectrum follow a power law $E^{-2.6}$, which is connected to the cascading process of the parent pion\cite{Cazon}. 
 Thus, a power law-like modification of the energy spectrum is introduced, with  the form:
\begin{equation}
f'_{X}(E_i)= f_{X}(E_i) \cdot E_{i}^{\delta} \:.
\label{eq: dist}
\end{equation}
This transformation is able to effectively modify the spectrum making it harder or softer, while maintaining the main characteristics essentially unchanged. The parameter $\delta$ is the number modulating the modification. In figure \ref{fig: MuSpectrum a}, the \epos proton is changed by $\delta=-0.1$ and $\delta=+0.1$, in green and black lines respectively. The corresponding ratios are displayed in figure \ref{fig: MuSpectrum b}. Note that the modification becomes a linear evolution as a function of $\log_{10} E$ as $E^{\delta} \simeq 1+\ln E$. It can be seen that \epos with $\delta=+0.1$ is very similar to \qgs iron, and thus $\delta=\pm0.1$ can be taken as the typical order of magnitude for the uncertainty of the energy spectrum of muons.

\begin{figure}[h]
\centering
\hspace{-0.5cm}
\begin{adjustwidth}{-0.50cm}{-0.50cm}
       \begin{subfigure}[b]{0.5\linewidth}\centering
\includegraphics[width=1\textwidth]{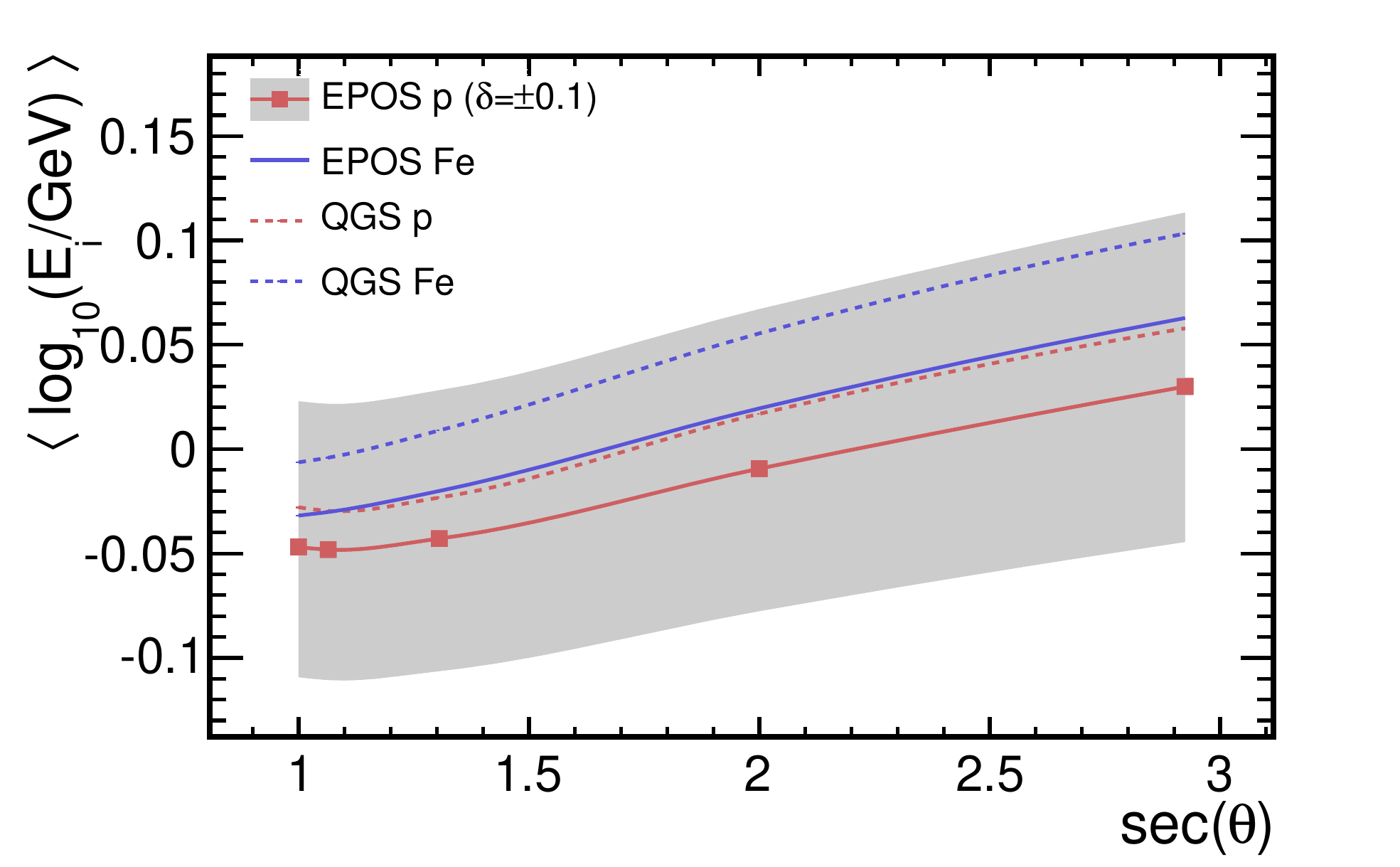}
                \caption{$\langle\log_{10}(E_{i}/\text{GeV})\rangle$ all muons at production}
                \label{fig: MuSpectrumTheta a}
        \end{subfigure}%
       \begin{subfigure}[b]{0.5\linewidth}\centering
\includegraphics[width=1\textwidth]{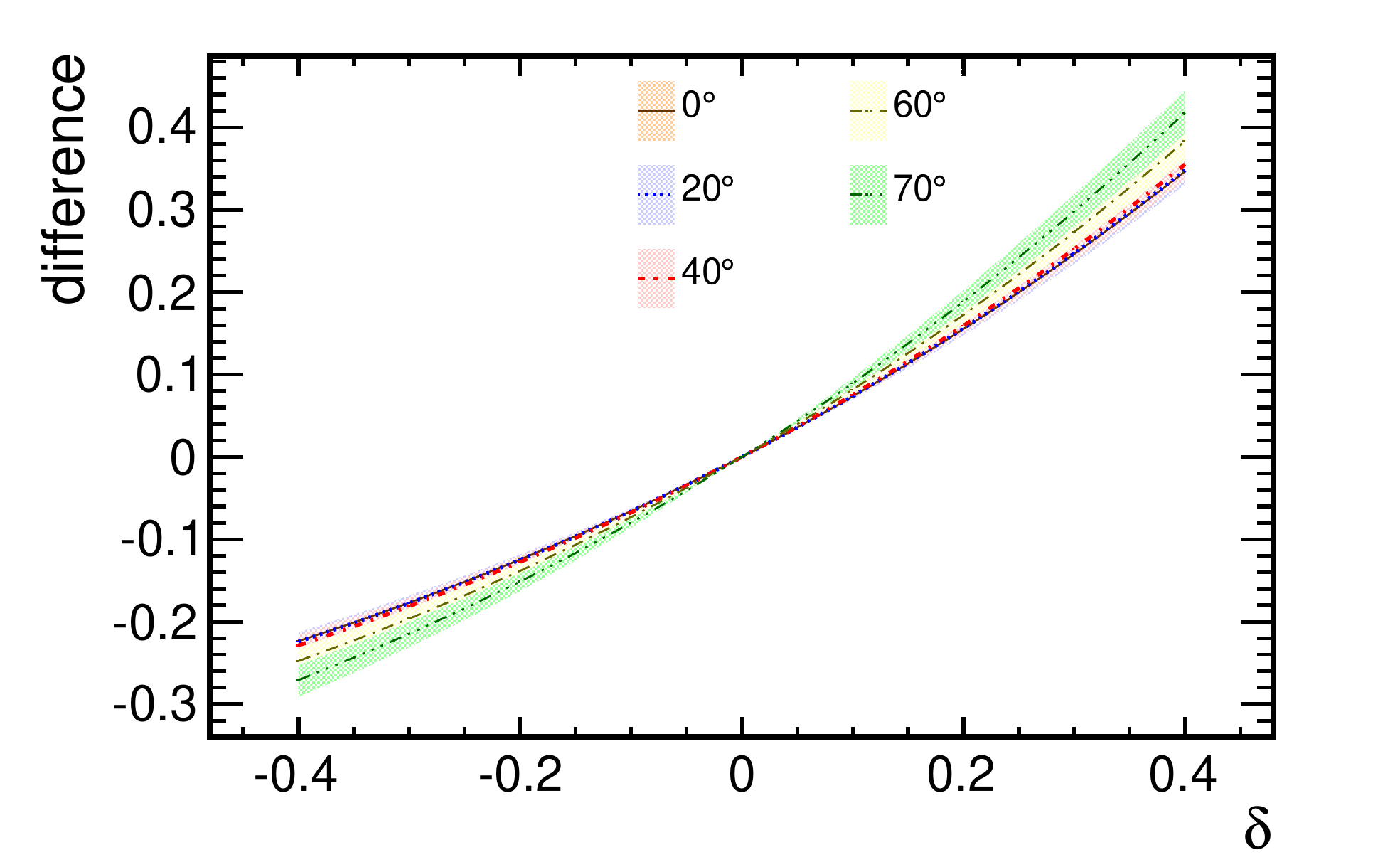}
                \caption{$\langle\log_{10}(E_{i}/\text{GeV})\rangle\DD{=\pm0.1}-\langle\log_{10}(E_{i}/\text{GeV})\rangle\DDz$}
                \label{fig: MuSpectrumTheta b}
        \end{subfigure}%
\end{adjustwidth}
\caption[]{Average energy at production ($\langle\log_{10}(E_{i}/\text{GeV})\rangle$) of all muons (a), and difference between the average logarithm of the energy (modified minus original), as a function of the $\delta$ parameter (b). The shadowed bands correspond to the different models and compositions, with the respective averages in dashed lines.
\label{fig: MuSpectrumTheta}}
\end{figure}

Figure \ref{fig: MuSpectrumTheta a}  displays the average production energy for both models and compositions. It also includes the average energy for proton \epos changed with $\delta=\pm0.1$.  In figure \ref{fig: MuSpectrumTheta b}, we can see the evolution of the average energy with the $\delta$ parameter for the modified proton \epos model (the shadow bands corresponds to the different models and composition).

\subsection{Fast transport model}
\label{subsection: Transport}

Full CORSIKA simulations are time and space consuming. Nevertheless, there is no need to repeat a full simulation each time the energy of muons at production is changed.
 Muons can be easily propagated to ground by means of a separate fast transport code that uses the modified distribution $F^\prime(X,E_i,cp_t)=F(X,E_i,cp_t)\cdot E_i^\delta$ as the only input (as demonstrated in \cite{CazonTransportModel}).

The transport code is implemented in two steps:  muons are produced along the shower axis at a depth $X$ (which corresponds to a distance from ground $z$), with an energy $E_i$, transverse momentum $p_t$ and a random polar angle, sampled from the $F(X, E_i, cp_t)$-distributions with weight $w_i$. The weight represent a bunch of muons and can be chosen to reduce the computation time. Muons are then propagated to ground following a straight line according to their 3-momentum, and accounting for the continuous energy loss; in a second step, the multiple scattering and magnetic field effects were included, the impact point on ground was corrected, and finally, the energy loss, arrival time delay, and decay probability are re-evaluated. At the end of the fast simulations, the distributions of muons at the ground is obtained as a function of: final weight $w_f$(due to the decay probability), distance to the shower axis $r$ and polar angle $\zeta$ of the impact point at ground, energy at ground $E_f$, and the different contributions to the overall arrival time delay with respect to a plane shower front, among them, the time delay due to subliminal velocities $t_\epsilon$, which is referred as {\it kinematic delay}  \cite{CazonTransportModel}.\\

In addition to energy loss processes, multiple scattering, and magnetic field deflections, muons can decay in flight with a certain probability, in which case some information is lost. The latter phenomenon plays a fundamental role in shaping the distributions of muons at the ground: the muon lateral distribution, time distribution and energy spectrum are suppressed in the regions dominated by low energy muons.

\begin{figure}[h]
\centering
\hspace{-0.5cm}
\begin{adjustwidth}{-0.50cm}{-0.50cm}
       \begin{subfigure}[b]{0.5\linewidth}\centering
\includegraphics[width=1\textwidth]{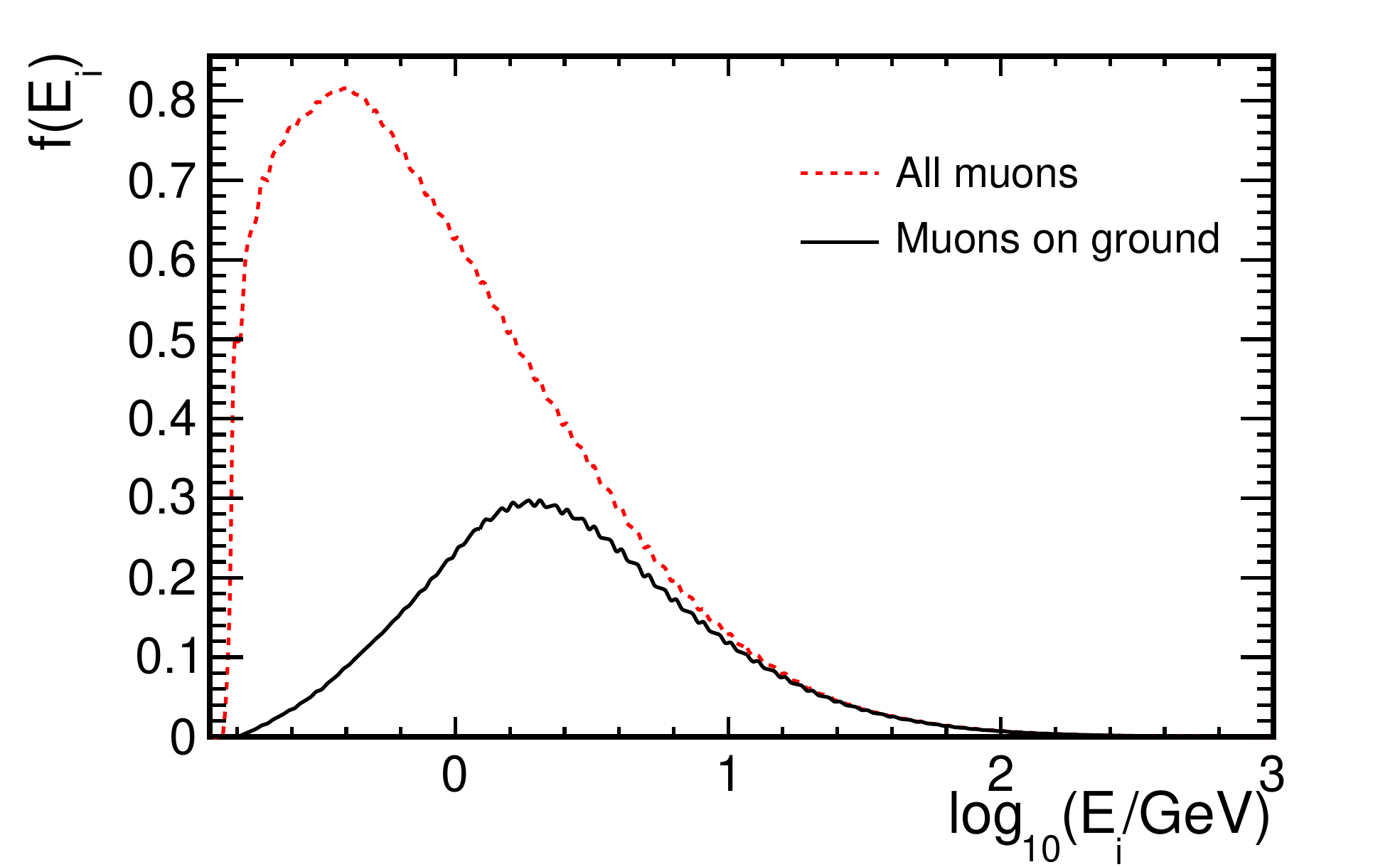}
                \caption{$\langle\log_{10}(E_{i}/\text{GeV})\rangle$ all muons }
                \label{fig: MuSpectrumAll a}
        \end{subfigure}%
       \begin{subfigure}[b]{0.5\linewidth}\centering
\includegraphics[width=1\textwidth]{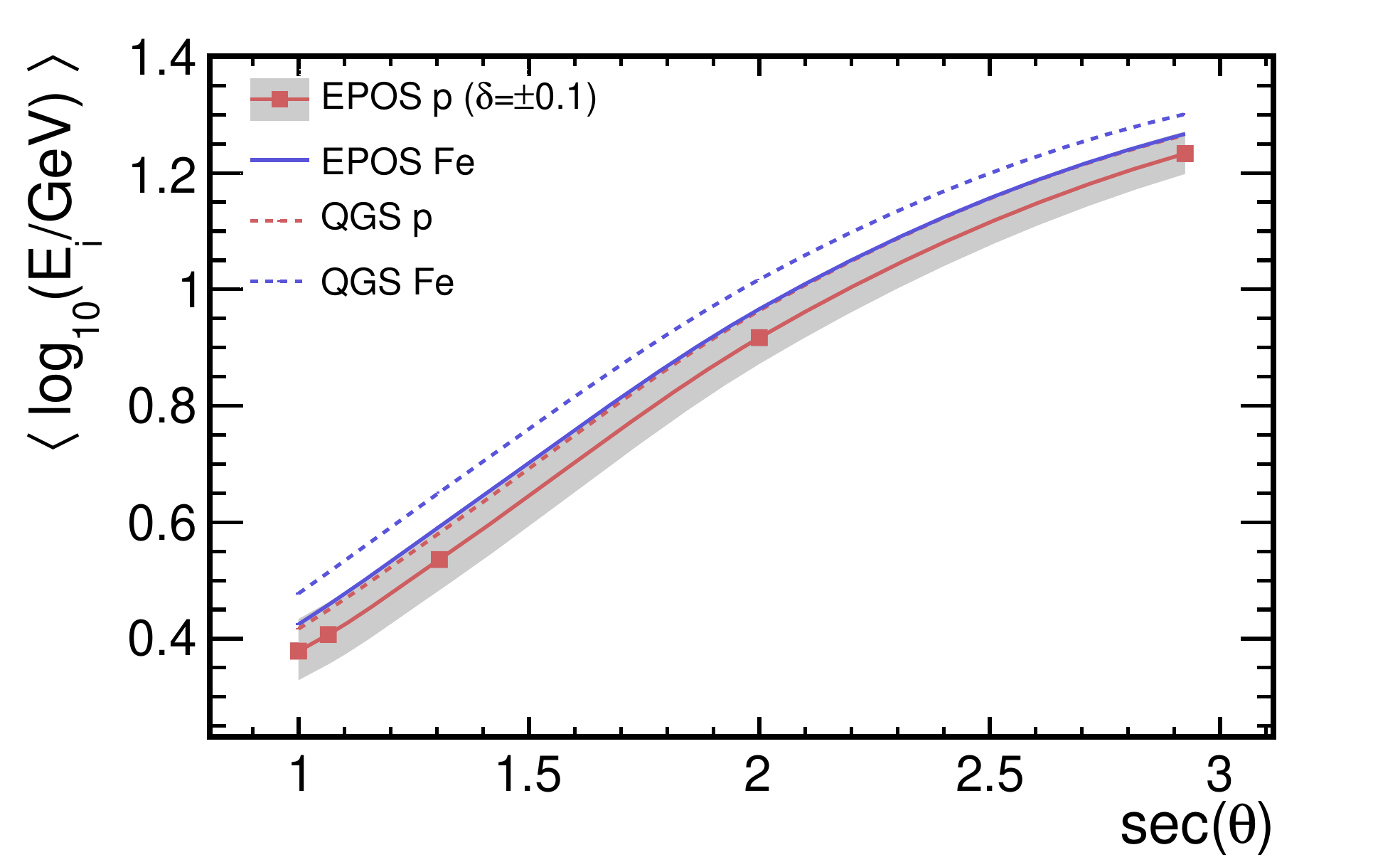}
                \caption{$\langle\log_{10}(E_{i}/\text{GeV})\rangle$ arriving at ground}
                \label{fig: MuSpectrumAll b}
        \end{subfigure}%
\end{adjustwidth}
\caption[]{Muon spectrum distribution of all produced muons (in dashed lines) and the ones arriving at the ground (full lines), for a simulation of proton showers using \epos model, at $20^{\circ}$ zenith angle, in (a). In (b), the average energy at production is plotted for muons arriving at ground.
\label{fig:MuSpectrumAll}}
\end{figure}

In figure \ref{fig: MuSpectrumAll a}, the dashed lines are the energy spectrum for all the produced muons for proton \epos while the full line represents the energy distributions of muons at production including only those that reach the ground. Most muons are produced below 3 GeV (peaking at  $\sim0.3$ GeV). A significant fraction of these low energy particles decay before reaching the ground while higher energy muons will mostly survive. 

From figure  \ref{fig: MuSpectrumAll b} it can be seen that at the ground the average $\langle\log_{10}(E_{i}/\text{GeV})\rangle$ for $\delta=+0.1$ with \epos proton is similar to iron \epos. This smaller effect, with respect to the effect of $\delta=+0.1$ at production, can be understood through atmospheric energy losses and muon decays.
%

\section{Number of muons on the ground}
\label{section: Number of muons}

\begin{figure}[h]
\centering
\hspace{-0.5cm}
\begin{adjustwidth}{-0.50cm}{-0.50cm}
       \begin{subfigure}[b]{0.5\linewidth}\centering
\includegraphics[width=1\textwidth]{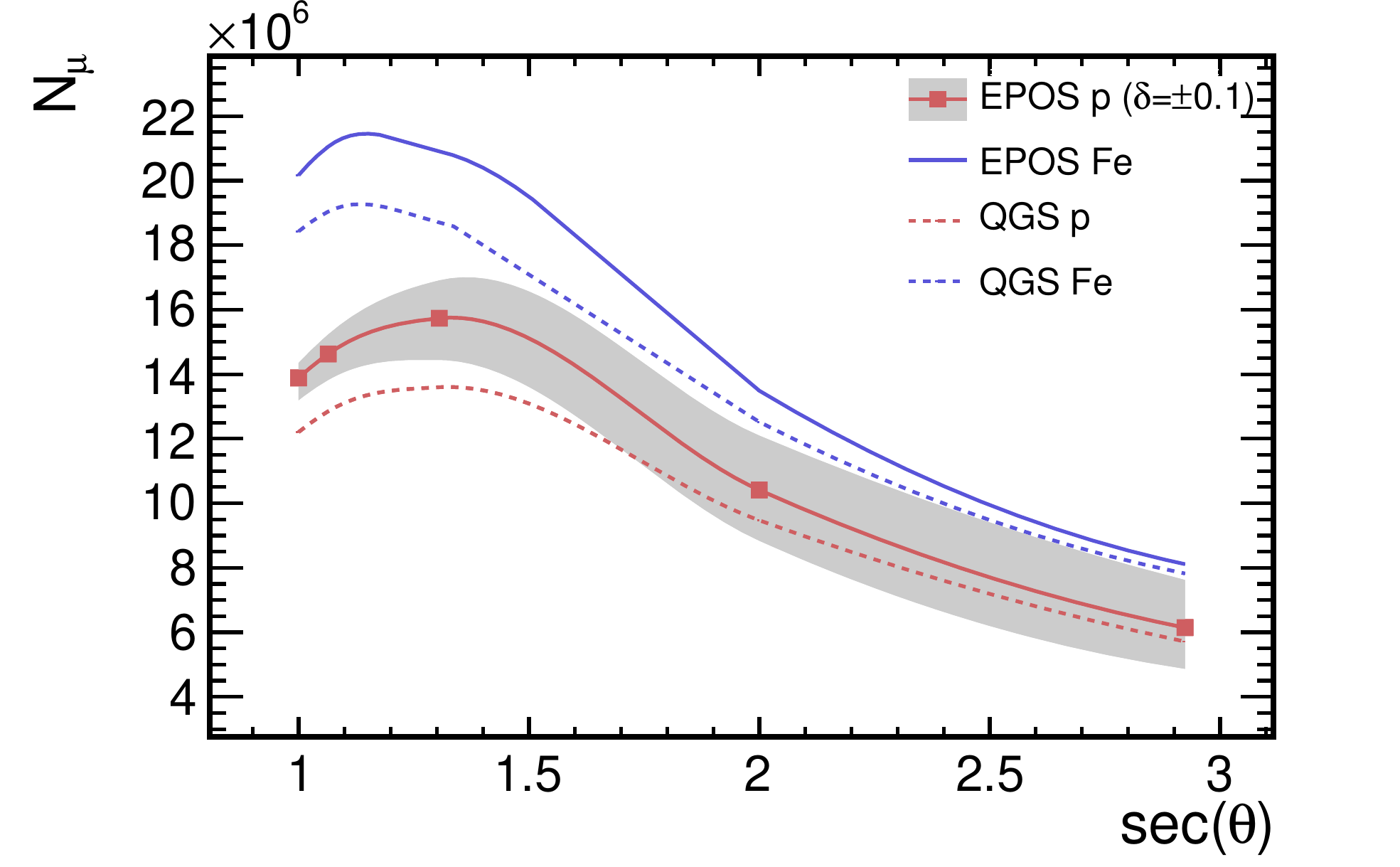}
                \caption{$N_{\mu}$ }
                \label{fig: MuNintegral a}
        \end{subfigure}%
       \begin{subfigure}[b]{0.5\linewidth}\centering
\includegraphics[width=1\textwidth]{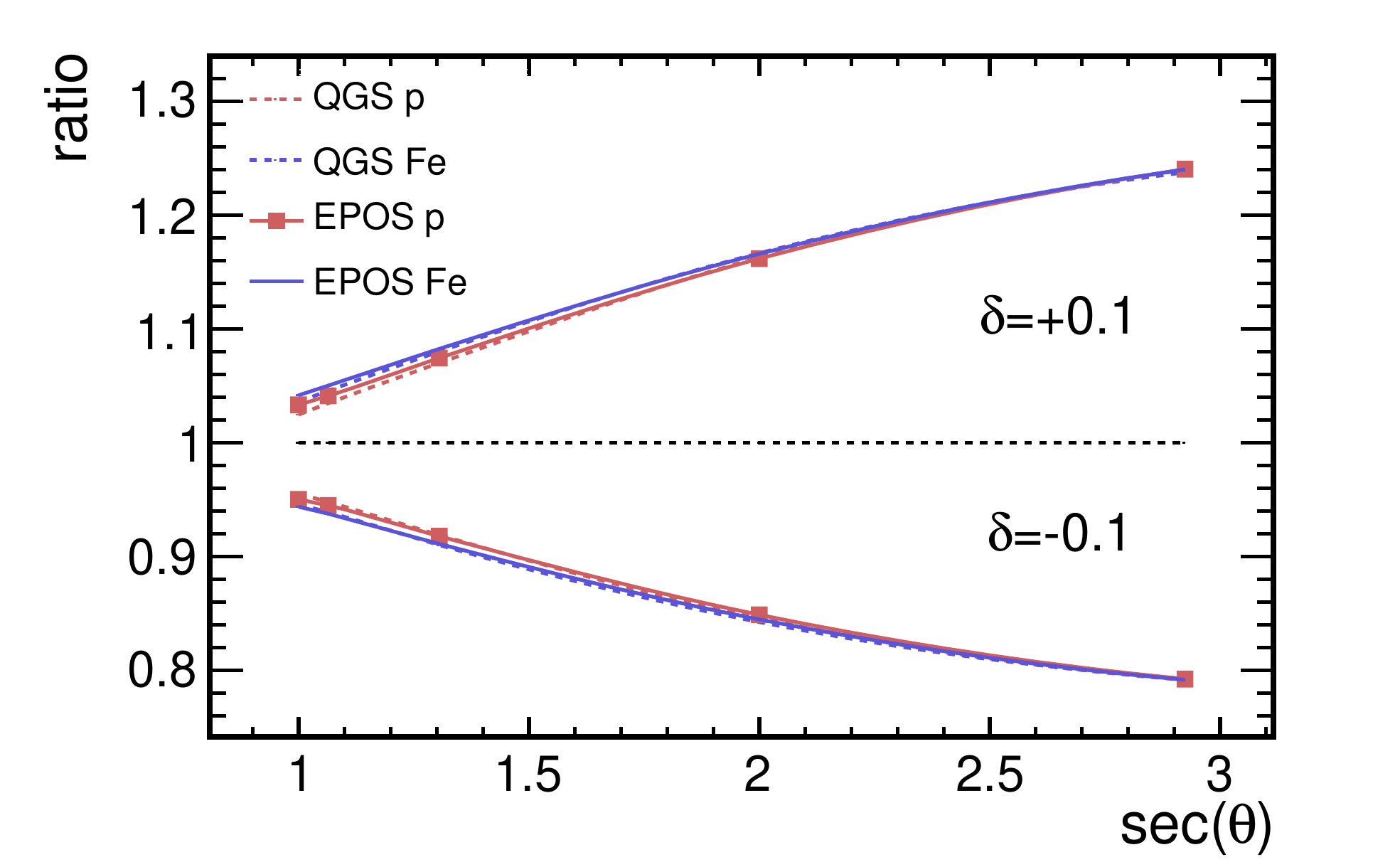}
                \caption{$N\DDsm{\mu}{=\pm0.1} / N\DDsm{\mu}{=0}$}
                \label{fig: MuNintegral b}
        \end{subfigure}%
\end{adjustwidth}
\caption[]{Number of muons arriving at ground between 500 m and 2000 m (a). Fraction of the number of muons arriving at the ground, between the changed muon spectrum and the initial spectrum, for the different models and compositions (b).
\label{fig: MuNintegral}}
\end{figure}

Near the shower core, cosmic ray detectors can be saturated, due to the high particle density. Far from the core,  the limited size of the detectors limits the sensitivity to very low particle densities (particle per unit area). Therefore, only muons between 500 m and 2000 m are considered, and a truncated number of muons is defined as  $N_\mu=\int \int_{500}^{2000} \rho_\mu(r) r dr d\zeta$. The number of muons at the ground is plotted in figure \ref{fig: MuNintegral a}. As expected, the iron samples and the \epos samples have more muons than proton or \qgs samples, respectively. 
From this figure it is also possible to see that a change in the proton \epos spectrum by $\delta=\pm0.1$ results in a modification of the number of muons at ground that still remains within the values for proton and iron with the \qgs model.\\
In figure \ref{fig: MuNintegral b}, the ratio of the muon number at ground between the modified and original spectrum distributions (ratio=$N_{\mu}(\delta)/N_{\mu}$) for each primary and model can be seen. Notice that results are similar for the same value of $\delta$ in all primary and model samples. Moreover, one important prediction, of the muon energy spectrum modification, is the behaviour with zenith angle. The relative importance of this effect increases with zenith angle. At higher zenith angles, muons will travel, on average, larger distances before reaching the ground, making the decay of low energy muons more important.\\

\begin{figure}[h]
\centering
\hspace{-0.5cm}
\begin{adjustwidth}{-0.50cm}{-0.50cm}
       \begin{subfigure}[b]{0.5\linewidth}\centering
\includegraphics[width=1\textwidth]{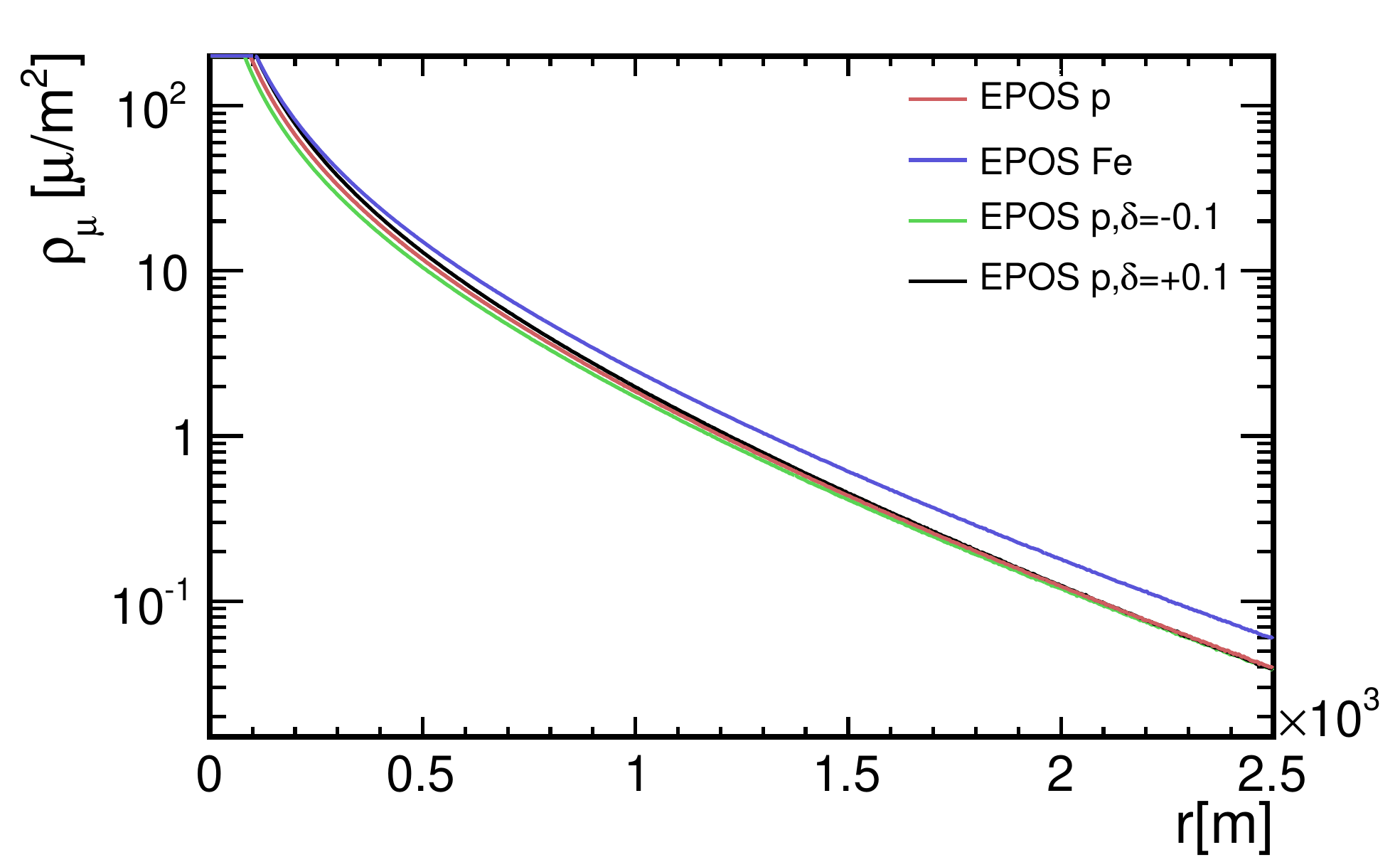}
                \caption{$\langle \rho_\mu (r)\rangle$ }
                \label{fig: LDF a}
        \end{subfigure}%
       \begin{subfigure}[b]{0.5\linewidth}\centering
\includegraphics[width=1\textwidth]{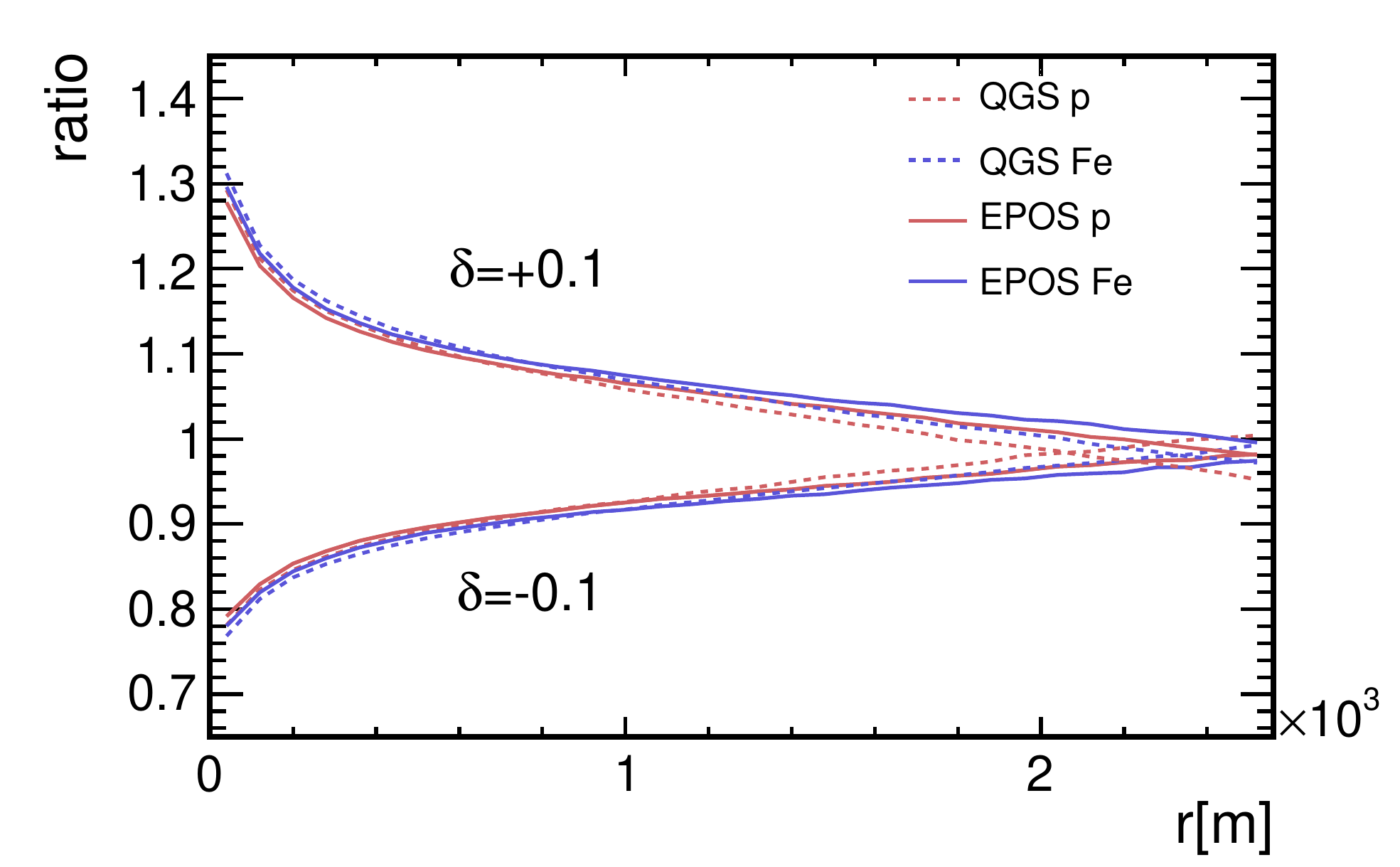}
                \caption{$\langle \rho_\mu (r)\rangle\DDsm{ }{=\pm0.1}/\langle \rho_\mu (r)\rangle\DDz $}
                \label{fig: LDF b}
        \end{subfigure}%
\end{adjustwidth}
\caption[]{Lateral muon distribution on ground at $40^\circ$ (a) and ratio between the changed (with $\delta=\pm0.1$) and unchanged LDF, for all models and compositions (b).
\label{fig: LDF}}
\end{figure}

The average \gls{ldf} of the muons over a given sample of showers with the same primary and zenith angle can be built as the average number of muons per unit area at ground as a function of the perpendicular distance to the shower axis $r$. Figure \ref{fig: LDF} shows an example for $40^\circ$ showers. From figure \ref{fig: LDF b} it is possible to see that variations $\delta=\pm0.1$ induces a change of the order of 10\% in the absolute muonic LDF. Moreover, it can be noted that the LDF is more sensitive to changes in the muon energy spectrum for lower distances to the shower core.

A modified NKG\cite{Nishimura} function 
\begin{equation}
f_{LDF}(r,\rho_{1000},\beta_{\mu}) =\rho_{1000}\left(\frac{r}{r_{opt}}\right)^{\beta_{\mu}}  \left(\frac{r+r_{scale}}{r_{opt}+r_{scale}}\right)^{\beta_{\mu}} \:,
\label{eq: LDF}
\end{equation}
with $r_{opt}=1000$ m and $r_{scale}=700$ m, was chosen to fit the average LDF, where $\rho_{1000}$ (normalization) and $\beta_{\mu}$ (LDF {\it slope}) were left as free parameters.  The range of distances to the core allowed in the fit ($r\in$[500,2000] m) justifies this particular choice of only one degree of freedom for the  shape parameter $\beta_{\mu}$.

\begin{figure}[h]
\centering
\hspace{-0.5cm}
\begin{adjustwidth}{-0.50cm}{-0.50cm}
       \begin{subfigure}[b]{0.5\linewidth}\centering
\includegraphics[width=1\textwidth]{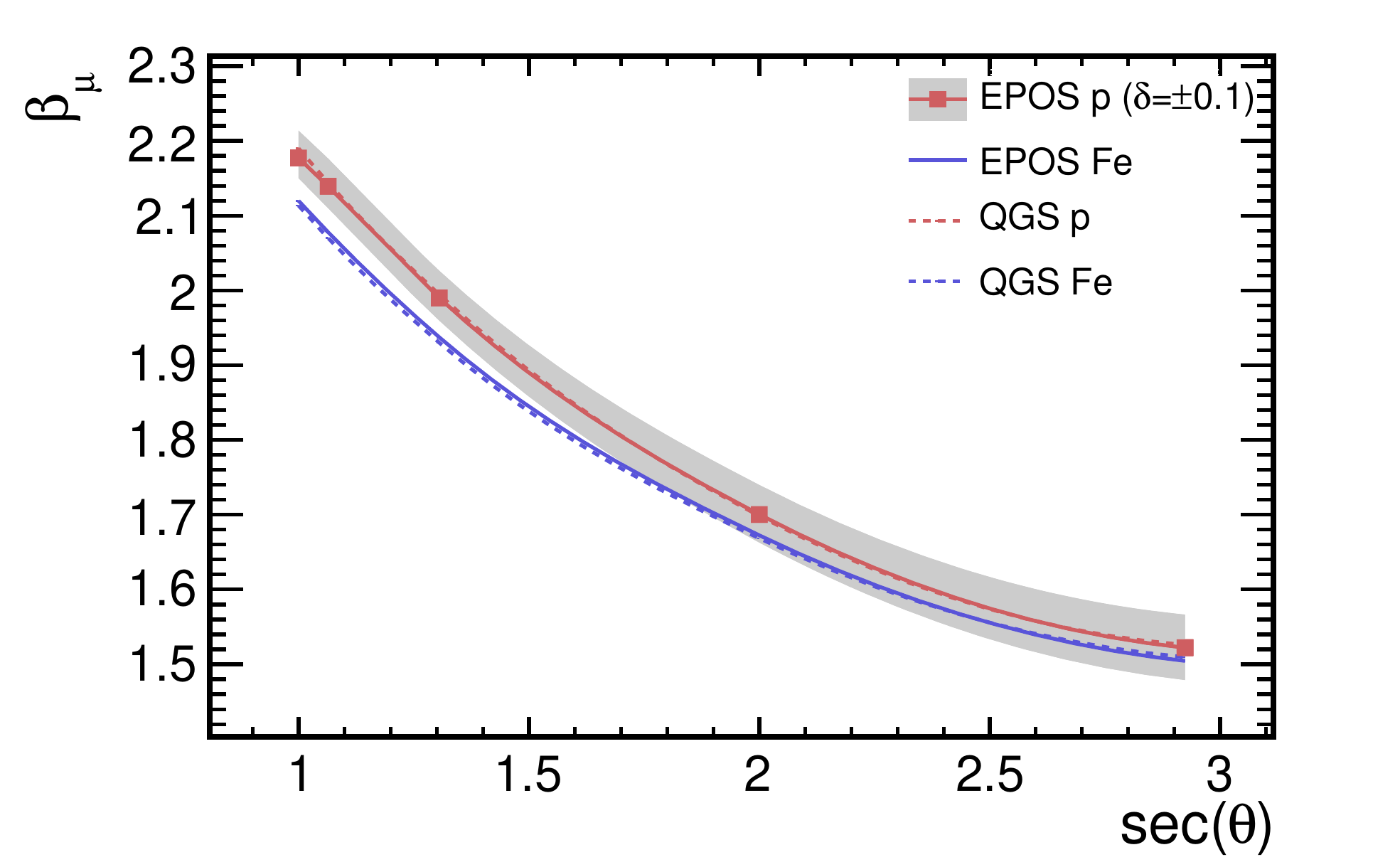}
                \caption{$\beta_{\mu}$ }
                \label{fig: MuRhoBeta a}
        \end{subfigure}%
       \begin{subfigure}[b]{0.5\linewidth}\centering
\includegraphics[width=1\textwidth]{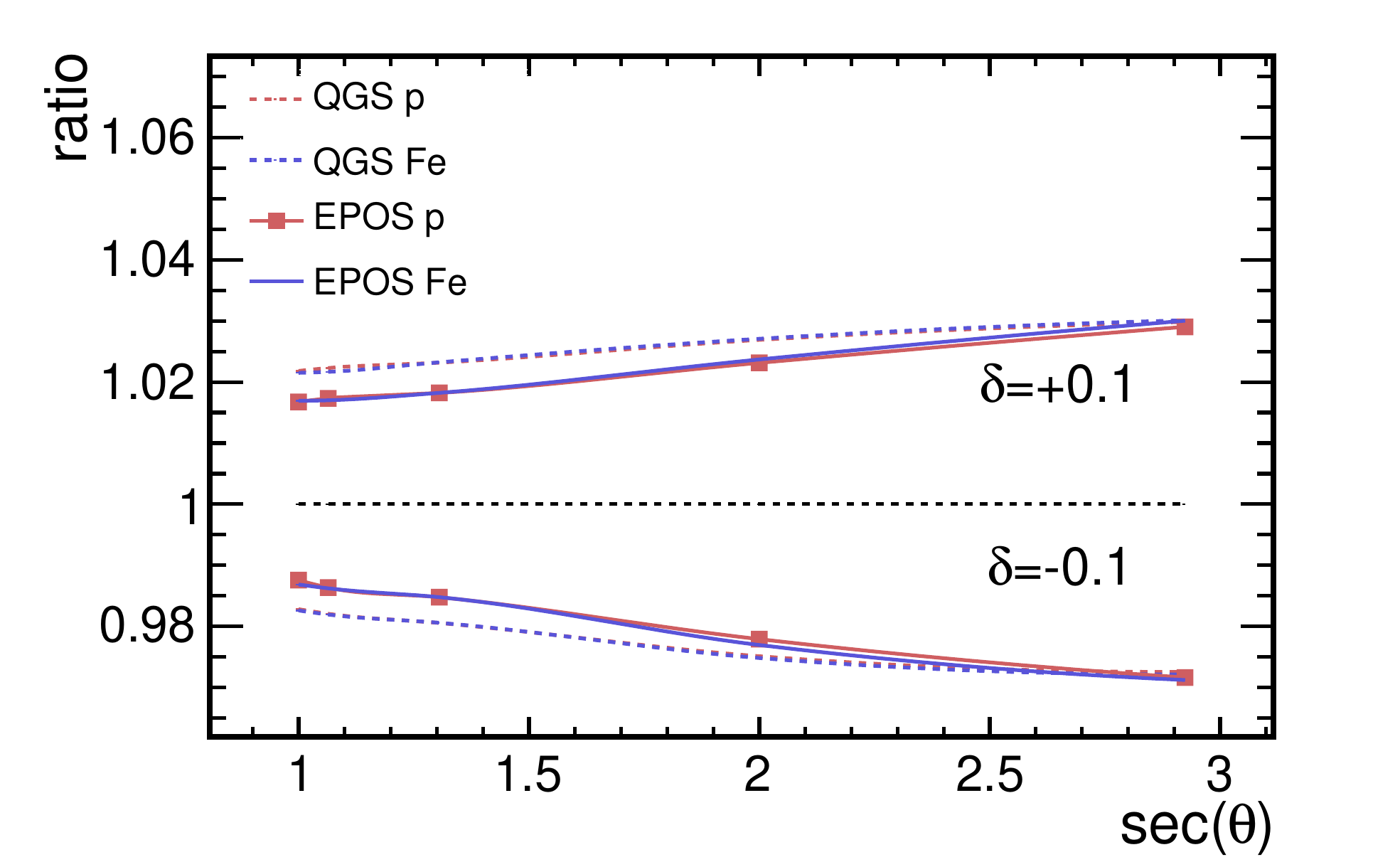}
                \caption{$\beta\DDsm{\mu}{=\pm0.1}/\beta\DDzMU{\mu}$}
                \label{fig: MuRhoBeta b}
        \end{subfigure}%
\end{adjustwidth}
\caption[]{The $\beta_{\mu}$ parameter of eq. \ref{eq: LDF}, fitted to the lateral distribution of muons on the ground (a). Comparison between the changed spectrum models and unchanged models, $\beta\DDsm{\mu}{=\pm0.1}/\beta\DDsm{\mu}{=0}$ (b).
\label{fig: MuRhoBeta}}
\end{figure}

The results for $\beta_{\mu}$ are plotted in figure \ref{fig: MuRhoBeta}. The slope is very similar in all samples, and the differences around $\sim5\%$ come essentially from the composition and not from the model. The effect of changing the muon spectrum is similar for all models and primaries and is around $\sim2\%$.

It is interesting to see that the $\beta_{\mu}$ parameter is not very sensitive to a change in the spectrum, while the total number of muons is more sensitive. More important is that the dependence of the muon number with zenith angle changes with the energy spectrum modification, which is a distinctive behaviour, while $\beta_{\mu}$ is constant.
These features could be used to identify discrepancies between data and simulations regarding the muon energy spectrum. \\


\FloatBarrier

\begin{figure}[!b]
\centering
\hspace{-0.5cm}
\begin{adjustwidth}{-0.50cm}{-0.50cm}
       \begin{subfigure}[b]{0.5\linewidth}\centering
\includegraphics[width=1\textwidth]{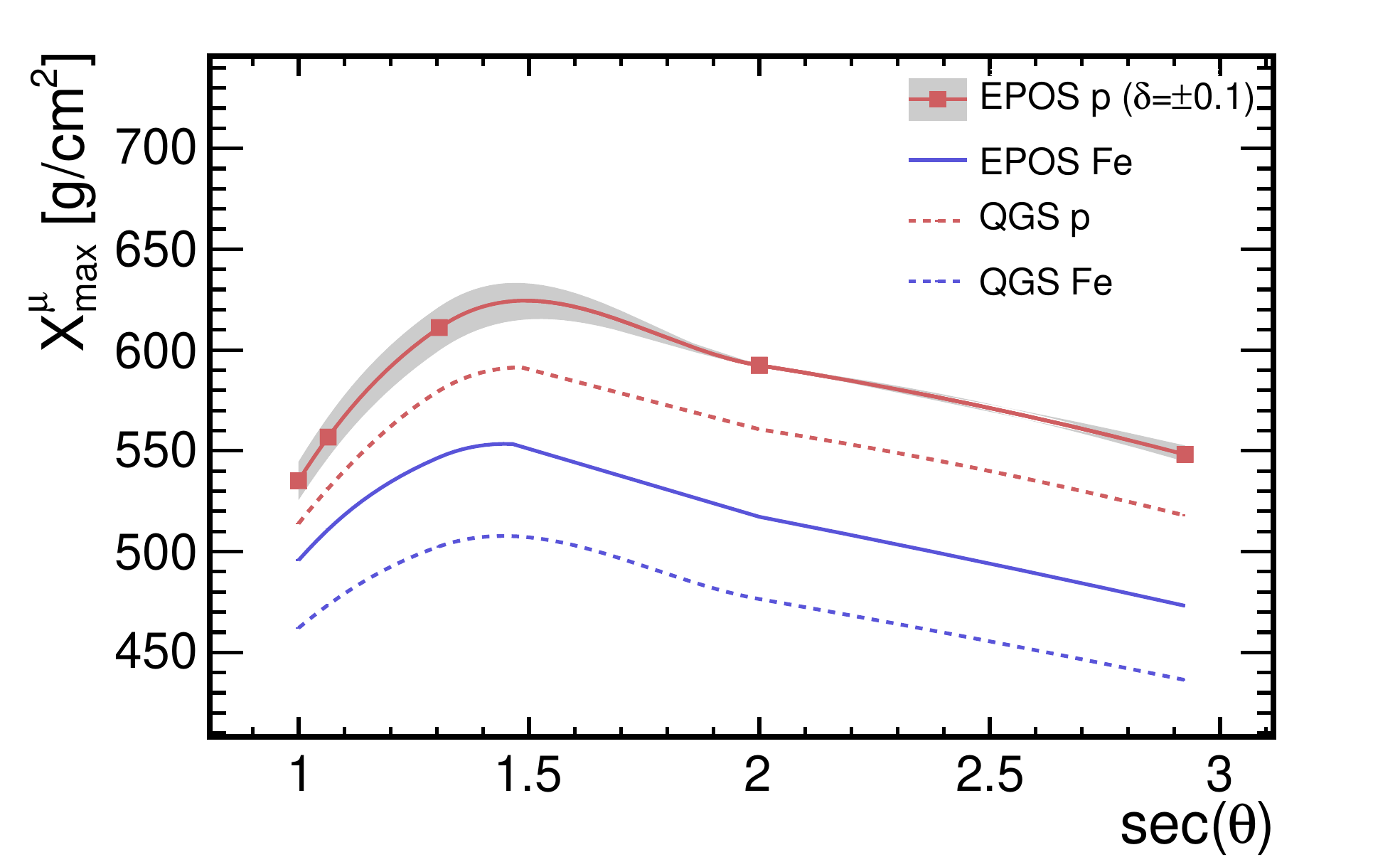}
                \caption{$\langle X_{max}^{\mu}\rangle$}
                \label{fig: MuXmaxApp a}
        \end{subfigure}%
       \begin{subfigure}[b]{0.5\linewidth}\centering
\includegraphics[width=1\textwidth]{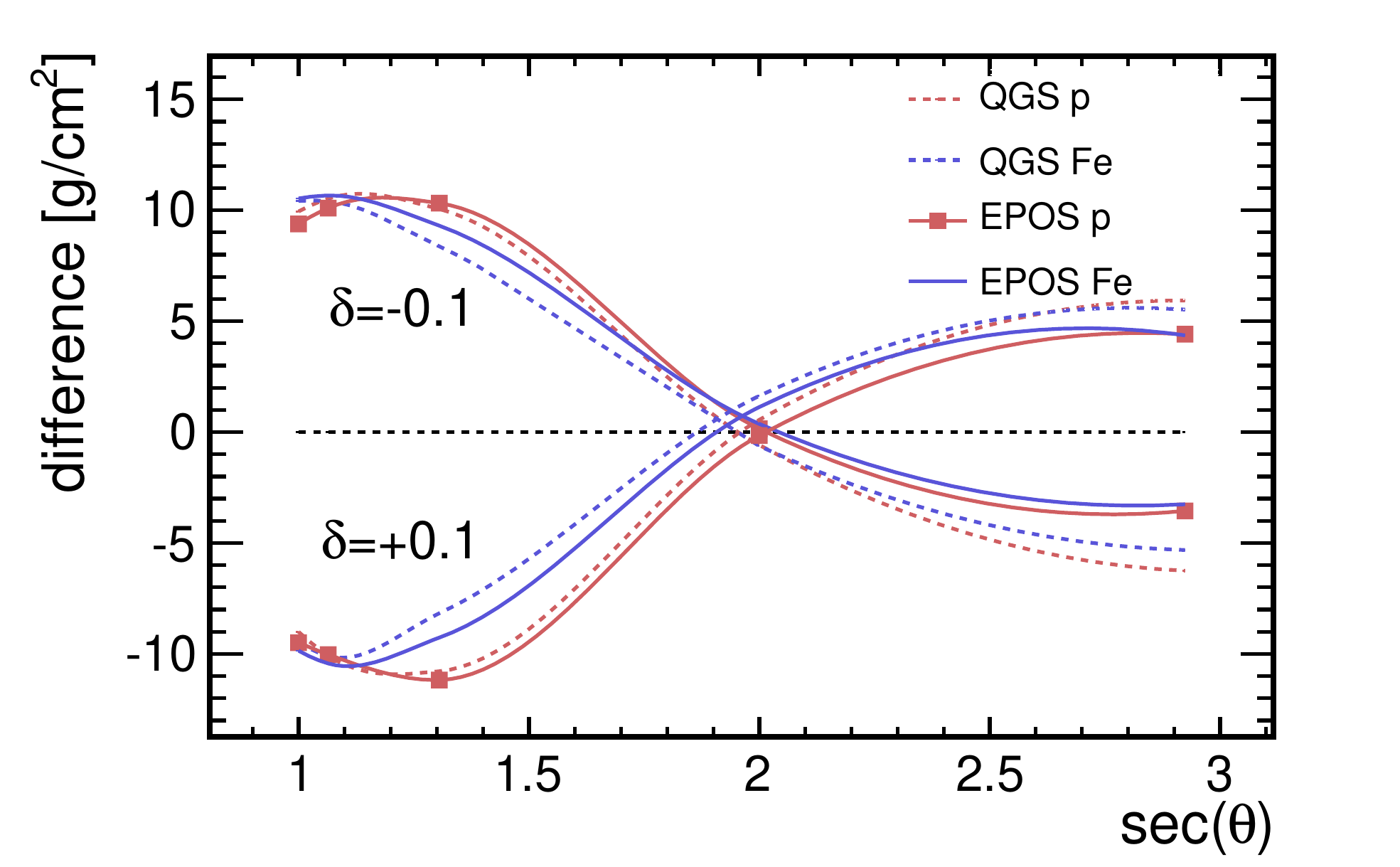}
                \caption{$\langle X_{max}^{\mu}\rangle\DD{=\pm0.1} - \langle X_{max}^{\mu}\rangle\DDz$}
                \label{fig: MuXmaxApp b}
        \end{subfigure}%
\end{adjustwidth}
\caption[]{Apparent $X_{max}^{\mu}$ seen between 1700 m and 4000 m (a). Effect of changing the spectrum between changed and unchanged models, as $\langle X_{max}^{\mu}\rangle\DD{=\pm0.1} - \langle X_{max}^{\mu}\rangle\DDz$ (b).
\label{fig: MuXmaxApp}}
\end{figure}

\subsection{Maximum of the muon production depth, $X_{max}^{\mu}$, and arrival time delay}
\label{section: ApparentXmax}


The apparent MPD-distribution is made from muons arriving at a particular region at the ground. Thus, only those muons that subtend an angle $\alpha$ compatible with the solid angle contained by the observation region to the production point will contribute to the apparent MPD-distribution. Notice that since $\sin \alpha=cp_t / E_i$, this automatically selects a fraction of the muons that  were produced. Moreover, decay and other propagation effects also modify the shape of the MPD-distribution depending on the energy at which muons were produced. Altogether, despite having the same  {\it total} MPD-distribution, $h(X)$, a change in the energy spectrum effectively changes the {\it apparent} MPD-distribution which is observed at ground.


The $X_{max}^\mu$ is plotted in figure \ref{fig: MuXmaxApp a} for the different studied samples. In this plot, only the muons with distances between 1700 m and 4000 m to the core were tracked\footnote{We have also verified that the relative changes in the range 1000-4000 m are practically the same with less than $\sim1.5 \text{ g/cm}^2$ difference}. In this range, there are several equivalent energy thresholds for muons due to different distances crossed in the atmosphere, but it is enough to study the overall differences. 
The maximum is reached higher for iron primaries and in the \epos model compared to the \qgs. 
The modification in the energy spectrum with $\delta=+0.1$ changes the apparent MPD approximately $-10 \text{ g/cm}^2$ at $20^\circ$ while there is no effect at $60^\circ$. The effect appears again for higher zenith angles but with an opposite sign with respect to vertical showers ($\sim+5\text{ g/cm}^2$ at $70^\circ$).
Since the average energy of muons at production decreases with the increase of the depth, the modification of the energy spectrum by a factor $E^\delta$ increases the importance of high energy muons, which are produced in early stages of the shower. This makes $X_{max}^\mu$ to become smaller. On the other hand, for very inclined showers, for instance those with $\theta =  70^\circ$, all low energy muons are highly suppressed due to decays, and therefore this effect is not present anymore and higher order effects begin to give non-negligible contributions. These effects are essentially related with the angle of emission of the muons.

Variations of the energy spectrum produce a net change in the angular dependence of the $X_{max}^\mu$ of the apparent MPD of the order of $X_{max}^\mu(60^\circ)-X_{max}^\mu(40^\circ)\sim 10 \text{ g/cm}^2$.





In \cite{CazonTransportModel,Cazon2004} it is shown that there is a one-to-one correspondence between the delay accumulated with respect to a shower front plane, the {\it geometrical delay}, $t_{g}\simeq \frac{1}{2}\frac{r^2}{z}$, and the production depth of each individual muon through its production distance to the ground $z$. Furthermore, this is what is used by the Auger Collaboration \cite{cazon2005} to experimentally reconstruct the apparent MPD distribution of single events\cite{MPD2014,Laura}. A direct calculation shows that 
the time delay changes by $\sim 5$ ns at 40 degrees and 1000 m from the shower core.

\begin{figure}[!h]
\centering
\hspace{-0.5cm}
\begin{adjustwidth}{-0.50cm}{-0.50cm}
       \begin{subfigure}[b]{0.5\linewidth}\centering
\includegraphics[width=1\textwidth]{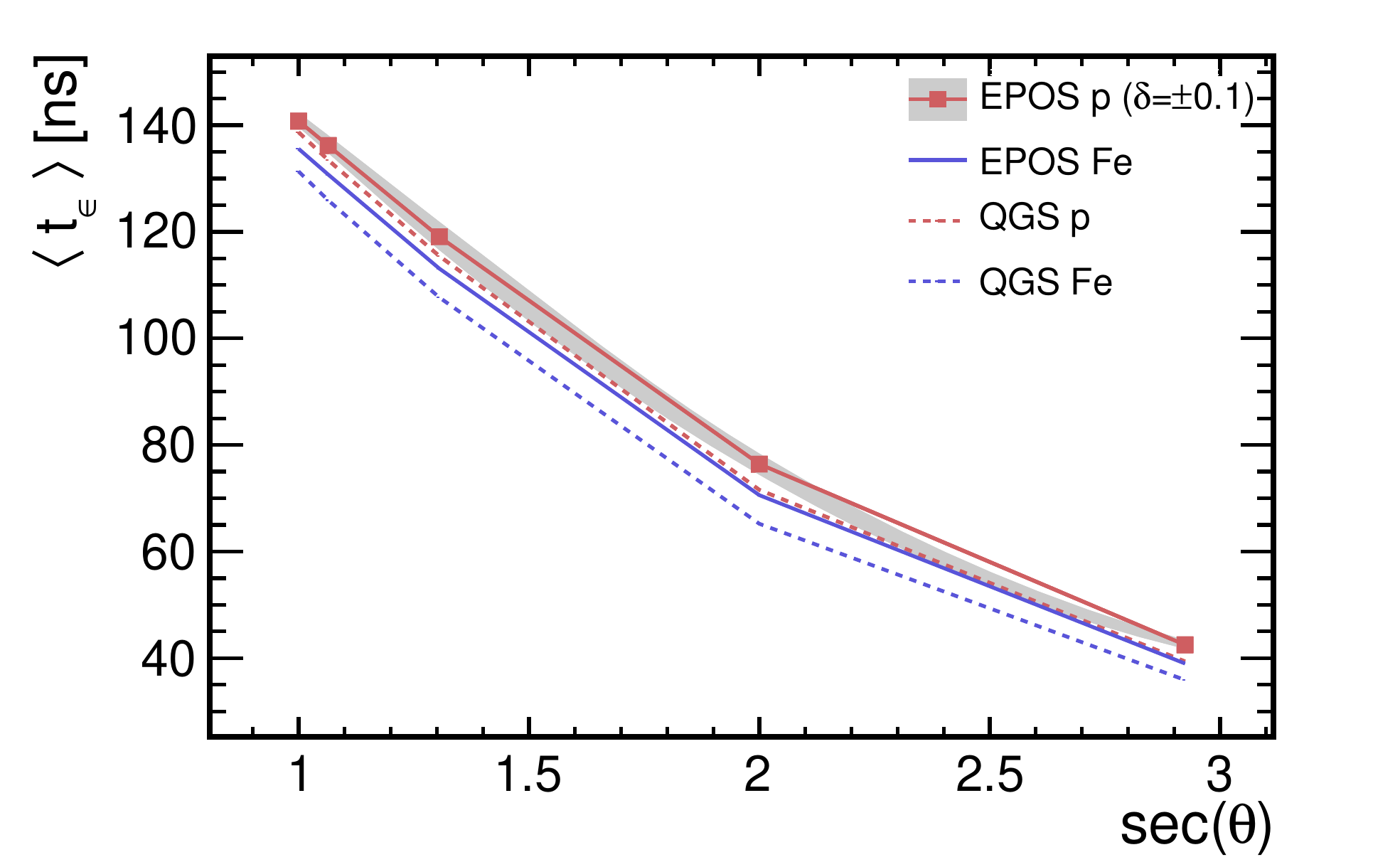}
                \caption{$\langle t_{\epsilon} \rangle$ at 1700 m}
                \label{fig: timeE 1700 a}
        \end{subfigure}%
       \begin{subfigure}[b]{0.5\linewidth}\centering
\includegraphics[width=1\textwidth]{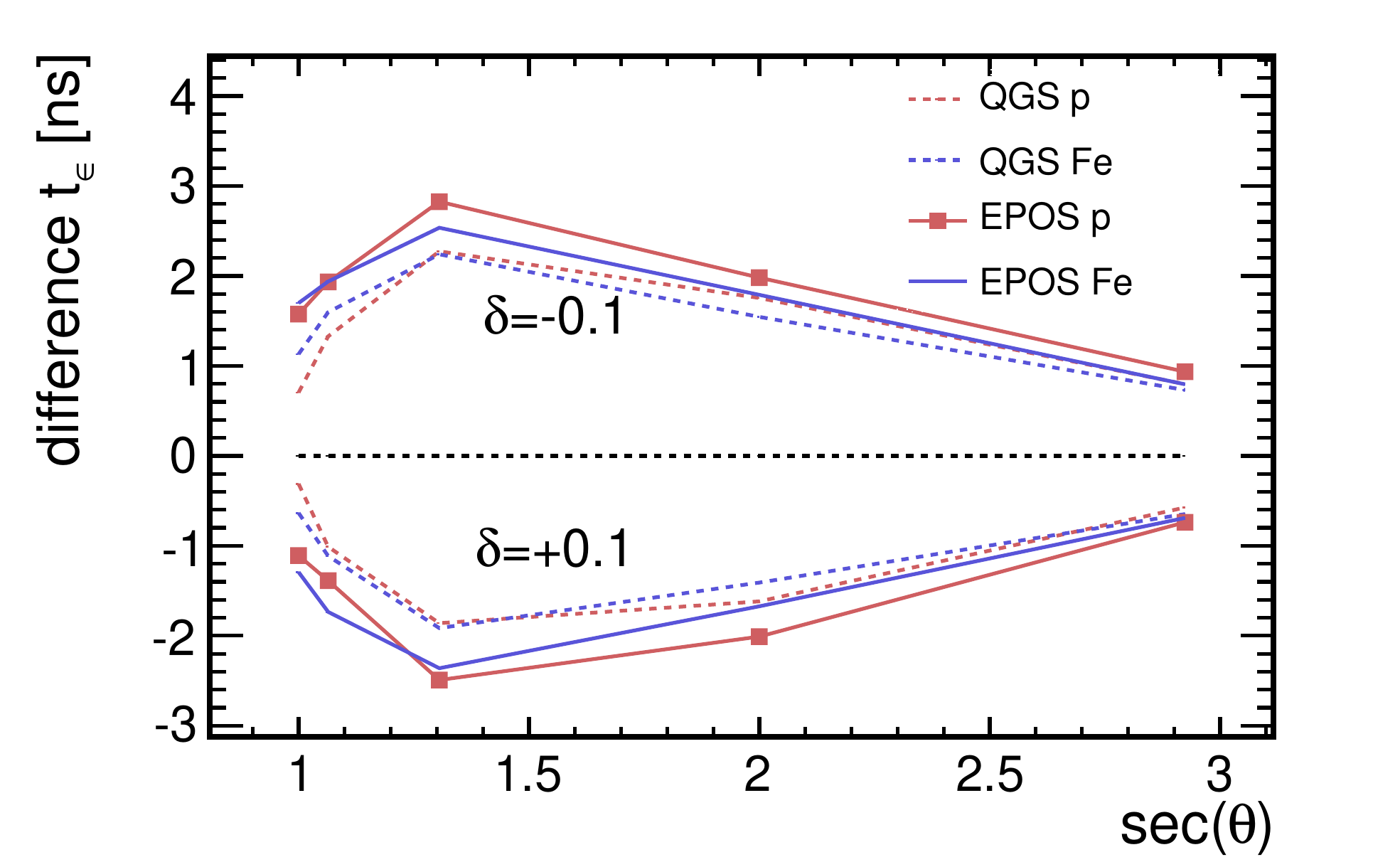}
                \caption{$\langle t_{\epsilon} \rangle\DD{=\pm0.1}  /  \langle t_{\epsilon}\rangle\DDz$}
                \label{fig: timeE 1700 b}
        \end{subfigure}%
\end{adjustwidth}
\caption[]{Kinematic delay time $t_{\epsilon}$ observed at 1700 m (a). Effect of changing the spectrum seen in $\langle t_{\epsilon}\rangle\DD{=\pm0.1} - \langle t_{\epsilon}\rangle\DDz$ (b) at 1700 m.\label{fig: timeE 1700}}
\end{figure}

The {\it kinematic delay}, $t_\epsilon$, acts as a second order correction to the total arrival time delay, and becomes less important as the distance to the shower core increases. Experimentally, it must be subtracted from the total time delay in order to access the geometric time delay. We have found that changes in the energy spectrum of muons produce a shift which is practically constant as a function of the distance to the shower core which is of the order of $\sim \pm$ 3 ns for $\delta=\mp0.1$ (at 1700 m, see figure \ref{fig: timeE 1700}). This would correspond to a bias on the reconstructed $X^\mu_{max}$ of the order of $\sim11 \text{ g/cm}^{2}$ at 60 degrees at $r=1000$ m or $\sim4 \text{ g/cm}^{2}$ at 1700 m.

\section{Signal in water Cherenkov detectors}
\label{section: Signal of muons}

In the previous section, the numbers of muons at ground were directly considered. Nevertheless, water Cherenkov detectors, like the ones used at the Pierre Auger Observatory or Ice Top, record signals with a non-negligible dependence on the energy of the muon. On one hand the emitted Cherenkov light depends on the velocity of the particle, until a plateau. On the other hand, high energy particles can also produce $\delta$-rays which generate an electromagnetic cascade inside of the tank giving additional contribution to the recorded signal. 

\begin{figure}[h]
\centering
\hspace{-0.5cm}
\begin{adjustwidth}{-0.50cm}{-0.50cm}
       \begin{subfigure}[b]{0.5\linewidth}\centering
\includegraphics[width=1\textwidth]{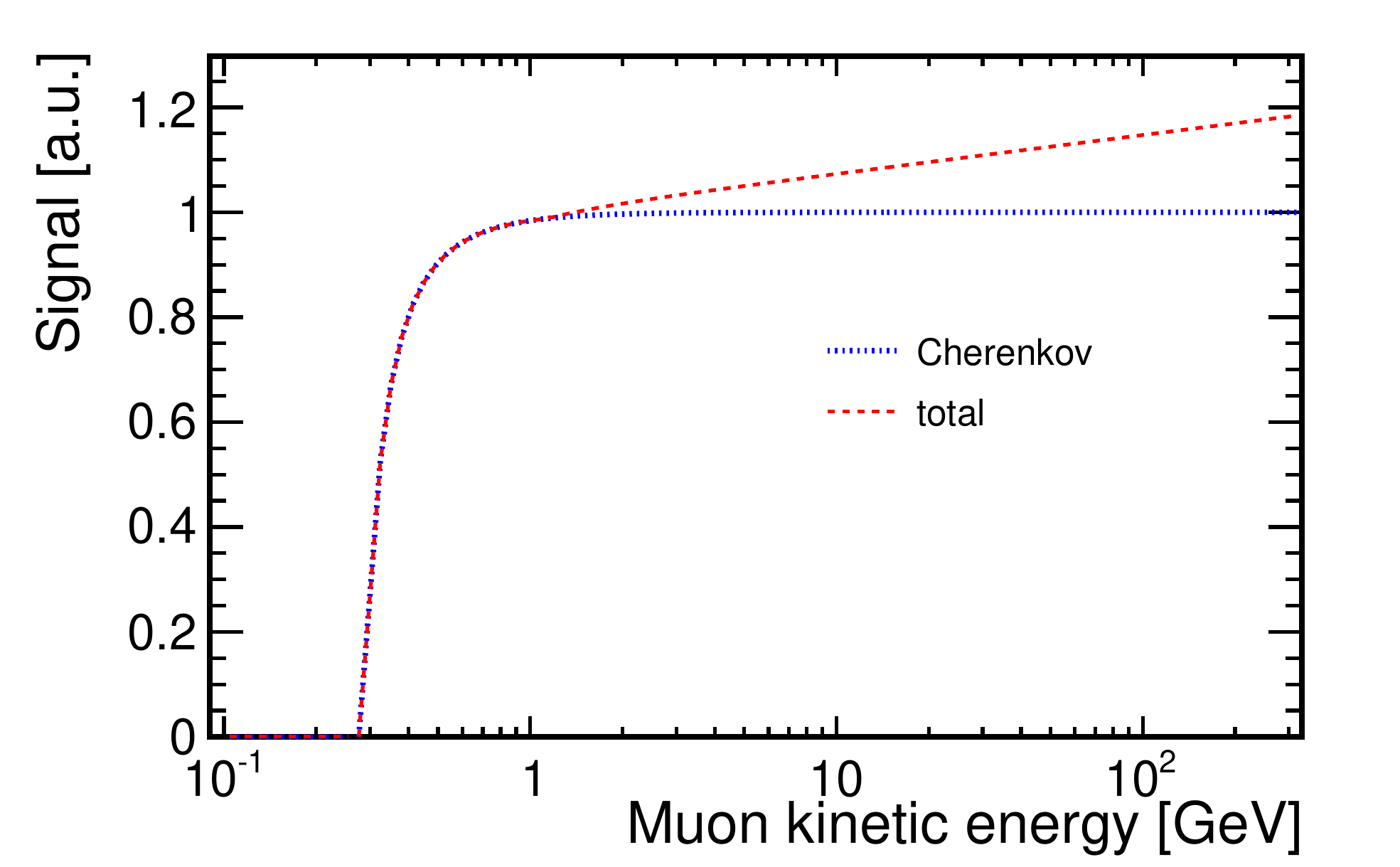}
                \caption{Muon $\langle Signal \rangle$ on water}
                \label{fig: MuEnergyGround a}
        \end{subfigure}%
       \begin{subfigure}[b]{0.5\linewidth}\centering
\includegraphics[width=1\textwidth]{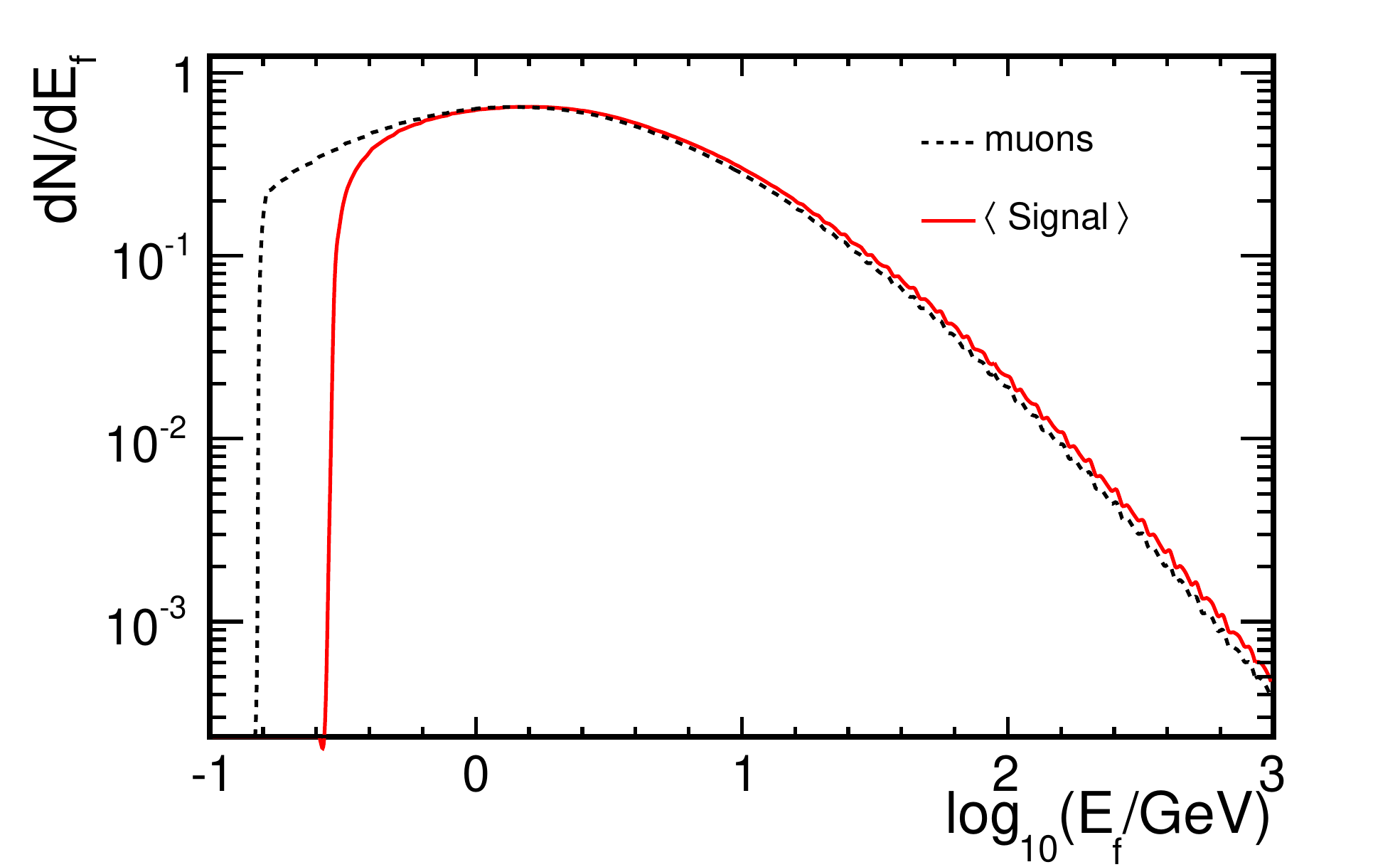}
                \caption{Energy distribution on ground}
                \label{fig: MuEnergyGround b}
        \end{subfigure}%
\end{adjustwidth}
\caption[]{Average signal of muons in water as function of the muon energy (a). Distribution of muons at the ground and weighted by the muon signal in dashed and full lines, respectively (b).
\label{fig: MuEnergyGround}}
\end{figure}

The average signal emitted by a muon is drawn in figure \ref{fig: MuEnergyGround a}.  The equation that translates the muon kinetic energy into signal (in arbitrary units) was obtained considering that the muon traverses vertically a volume of water with $1.2$ m thickness (similar to what happens for the Pierre Auger detectors). Two physical phenomena were considered: Cherenkov light emission and $\delta$-ray production. The first one has an explicit dependence on the muon velocity, $\beta=v/c$ and is represented in the figure by the blue dotted line. During the passage through the water the muon will lose on average $240\,$ MeV. This amount of energy was removed from the muon kinetic energy before computing the Cherenkov yield. Note that once the muon becomes \emph{relativistic} the number of emitted Cherenkov photons becomes independent of the energy of the particle. The $\delta$-ray effect was included following a parametrisation motivated from \cite{deltaRay}. For muons, this effect becomes more noticeable above $1\,$GeV and the contribution to the recorded signal increases logarithmically its energy.

In this section we calculate what would be the sensitivity to the changes on the energy spectrum of muons at production, of the signal observed by detectors similar to those used in the Auger Observatory.

In figure \ref{fig: MuEnergyGround b}, the muon energy at ground is shown in dashed lines and in full lines the same distribution weighted by the signal that muons leave in the water  is drawn (from figure \ref{fig: MuEnergyGround a}). The less energetic muons do not leave signal, while the higher energy tail is increased. 

\begin{figure}[h]
\centering
\hspace{-0.5cm}
\begin{adjustwidth}{-0.50cm}{-0.50cm}
       \begin{subfigure}[b]{0.5\linewidth}\centering
\includegraphics[width=1\textwidth]{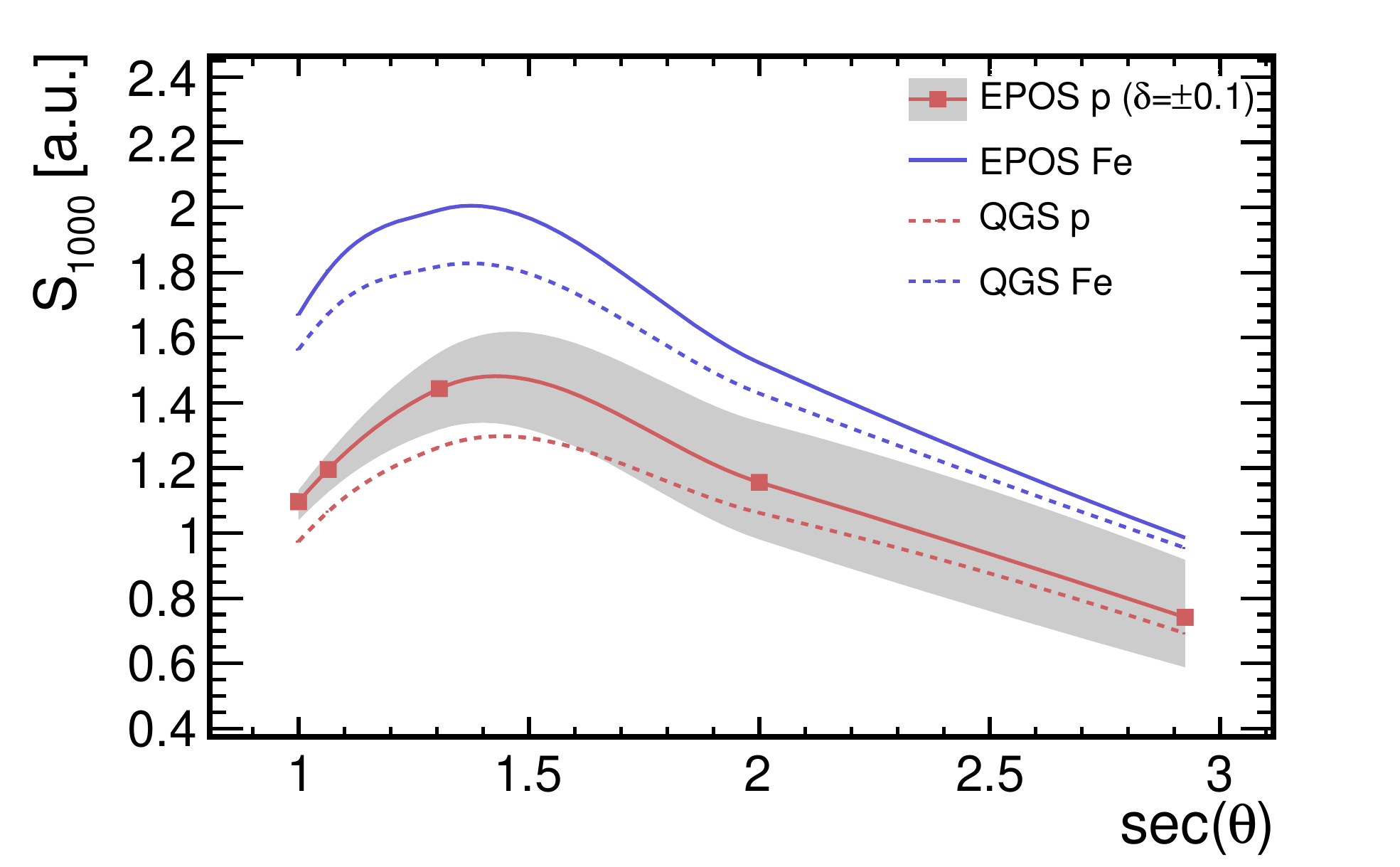}
                \caption{$S_{1000}$ }
                \label{fig: MuS1000 a}
        \end{subfigure}%
       \begin{subfigure}[b]{0.5\linewidth}\centering
\includegraphics[width=1\textwidth]{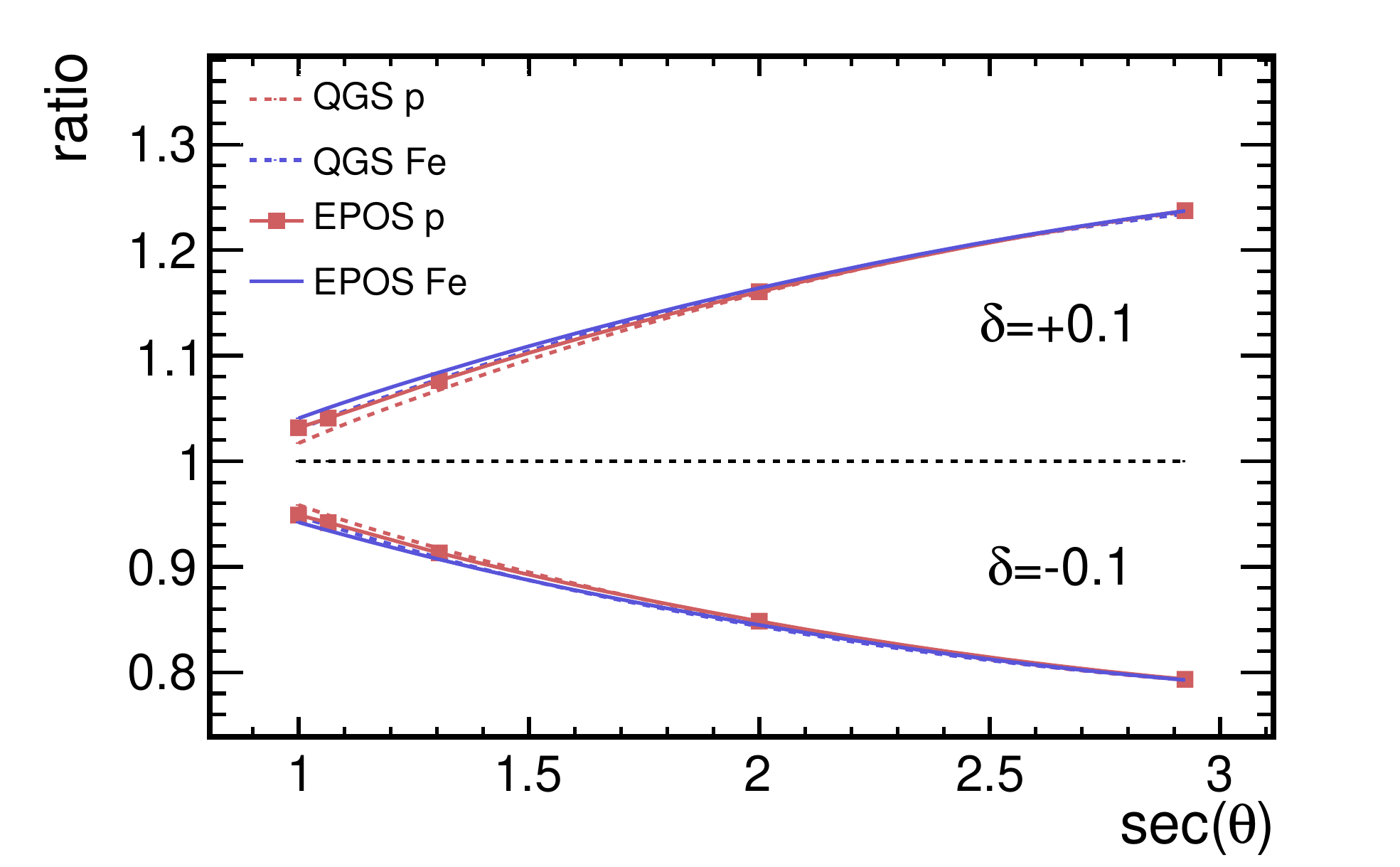}
                \caption{$S\DDsm{1000}{=\pm0.1}/S\DDzMU{1000} $ }
                \label{fig: MuS1000 b}
        \end{subfigure}%
\end{adjustwidth}
\caption[]{The $S_{1000}$ parameter of eq. \ref{eq: LDF} for the signals at the ground (a). Comparison between the changed spectrum models and unchanged models, $S\DDsm{1000,\mu}{=\pm0.1}/S\DDsmZ{1000,\mu}$ (b).
\label{fig: MuS1000}}
\end{figure}

The size parameter $S_{1000}$ (equivalent to $\rho_{1000}$), which is proportional to the total number of muons is displayed in figure \ref{fig: MuS1000 a}. Also here a very similar behaviour to that of the total number of muons (from figure \ref{fig: MuNintegral}) is observed.

\begin{figure}[h]
\centering
\hspace{-0.5cm}
\begin{adjustwidth}{-0.50cm}{-0.50cm}
       \begin{subfigure}[b]{0.5\linewidth}\centering
\includegraphics[width=1\textwidth]{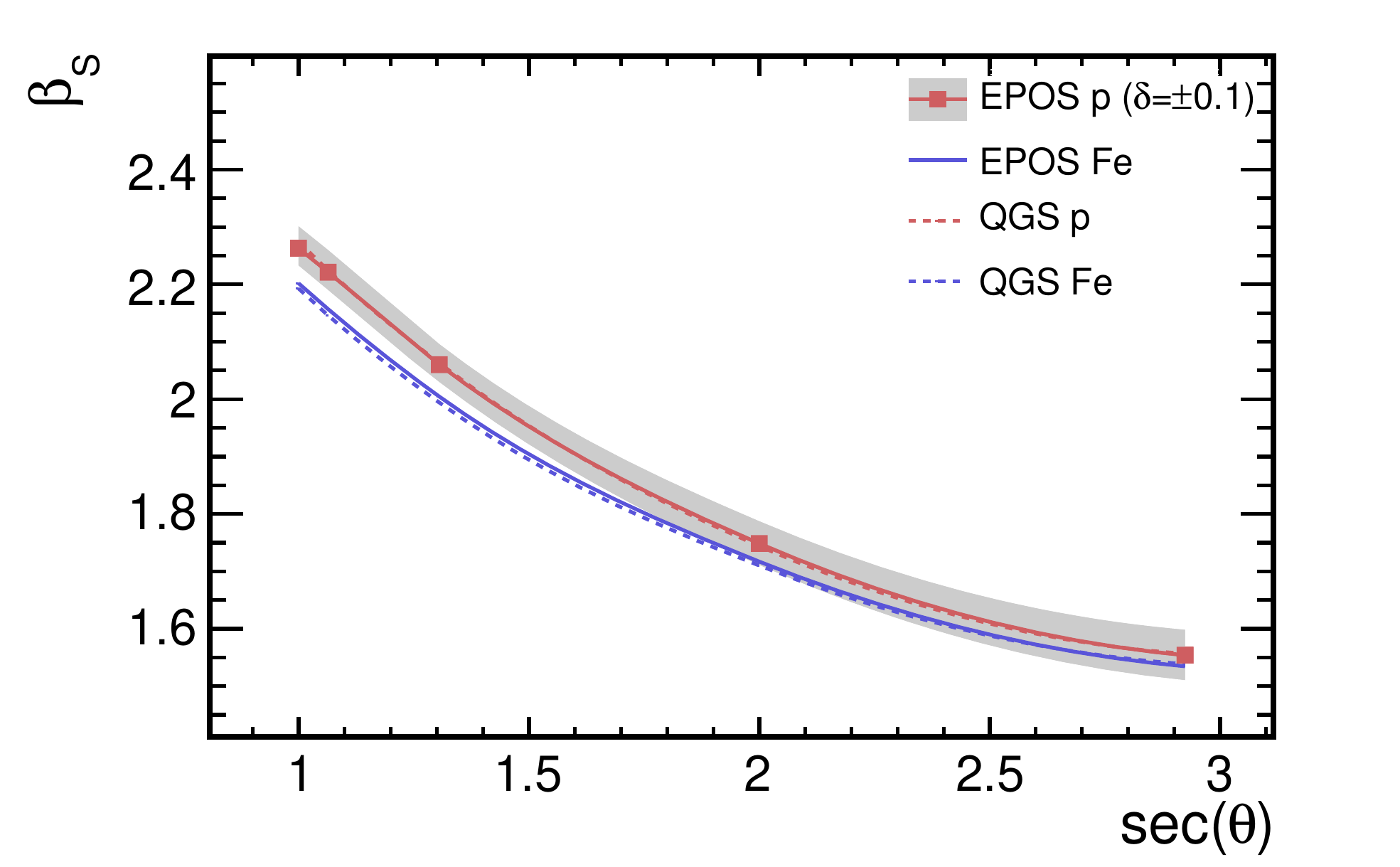}
                \caption{$\beta_{S}$ }
                \label{fig: MuBeta a}
        \end{subfigure}%
       \begin{subfigure}[b]{0.5\linewidth}\centering
\includegraphics[width=1\textwidth]{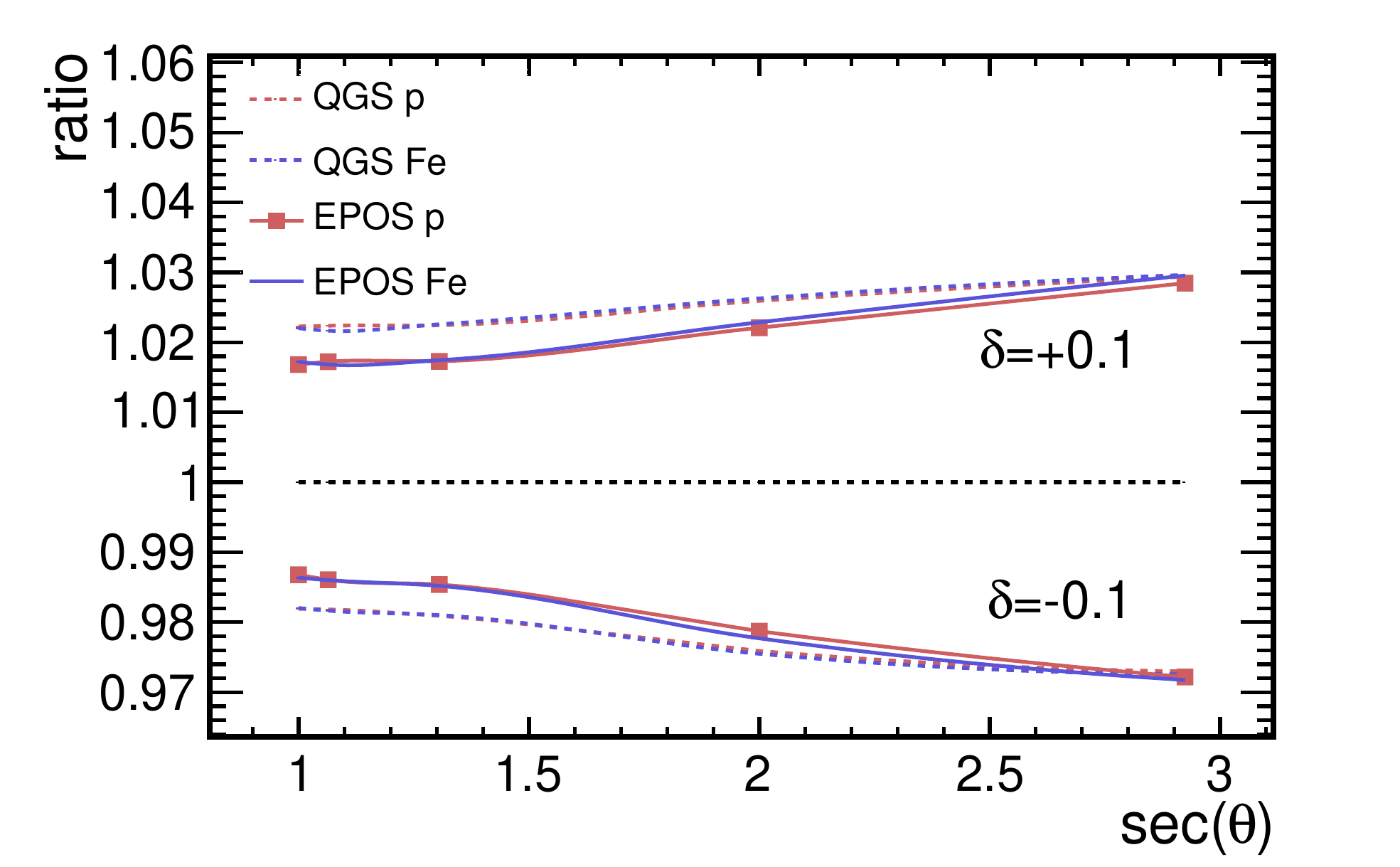}
                \caption{$\beta\DDsm{S}{=\pm0.1}/\beta\DDzMU{S}$}
                \label{fig: MuBeta b}
        \end{subfigure}%
\end{adjustwidth}
\caption[]{The $\beta_{S}$ parameter of eq. \ref{eq: LDF} for the signals on ground (a).Comparison between the changed spectrum models and unchanged models, $\beta\DDsm{S}{=\pm0.1}/\beta\DDsmZ{S}$ (b).
\label{fig: MuBeta}}
\end{figure}

The average \gls{ldf}, similar to figure \ref{fig: LDF}, can be built for the muonic signal at the ground and fitted again using eq. \ref{eq: LDF}. The slope $\beta_S$ and the size parameter $S_{1000}$ are plotted in figures \ref{fig: MuBeta} and \ref{fig: MuS1000}, respectively. 
The effect of changing the energy spectrum by $\delta$, on the parameter $\beta_S$ is similar to the one found in the previous section.

In summary, despite the dependence of the signal recorded by the water Cherenkov detectors with energy, all the tested observables present sensitivities to the muon energy spectrum that are similar to those found by considering directly the muons at the ground.

\FloatBarrier
\section{Sensitivity of the variables to the $\delta$ modification}
\label{section: Sensitivity}
\begin{figure}[h]
\centering
\hspace{-0.5cm}
\begin{adjustwidth}{-0.50cm}{-0.50cm}
       \begin{subfigure}[b]{0.5\linewidth}\centering
\includegraphics[width=1\textwidth]{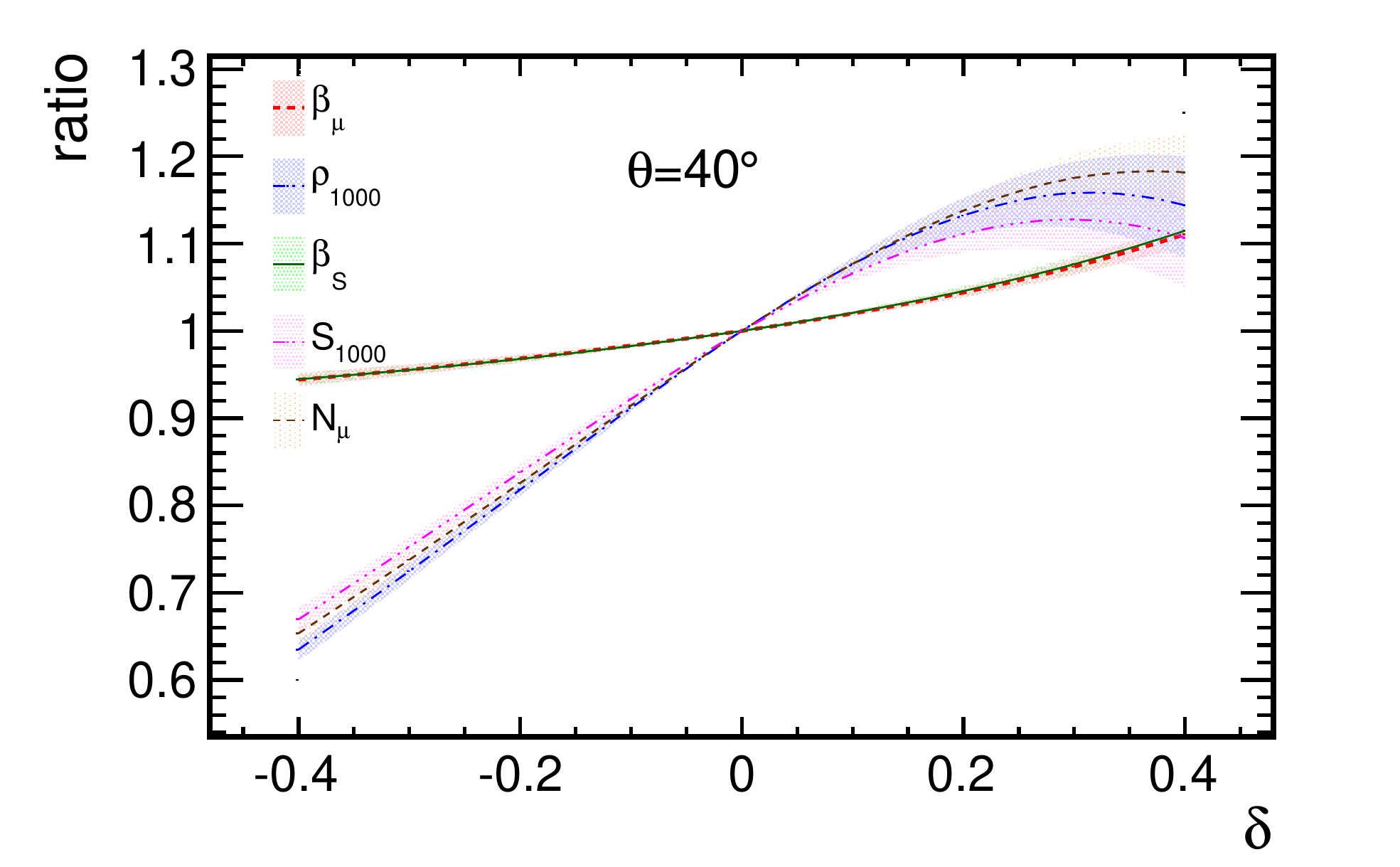}
                \caption{ratio $\frac{x\DD{ }}{x\DDz}$ }
                \label{fig: DELTA1 a}
        \end{subfigure}%
       \begin{subfigure}[b]{0.5\linewidth}\centering
\includegraphics[width=1\textwidth]{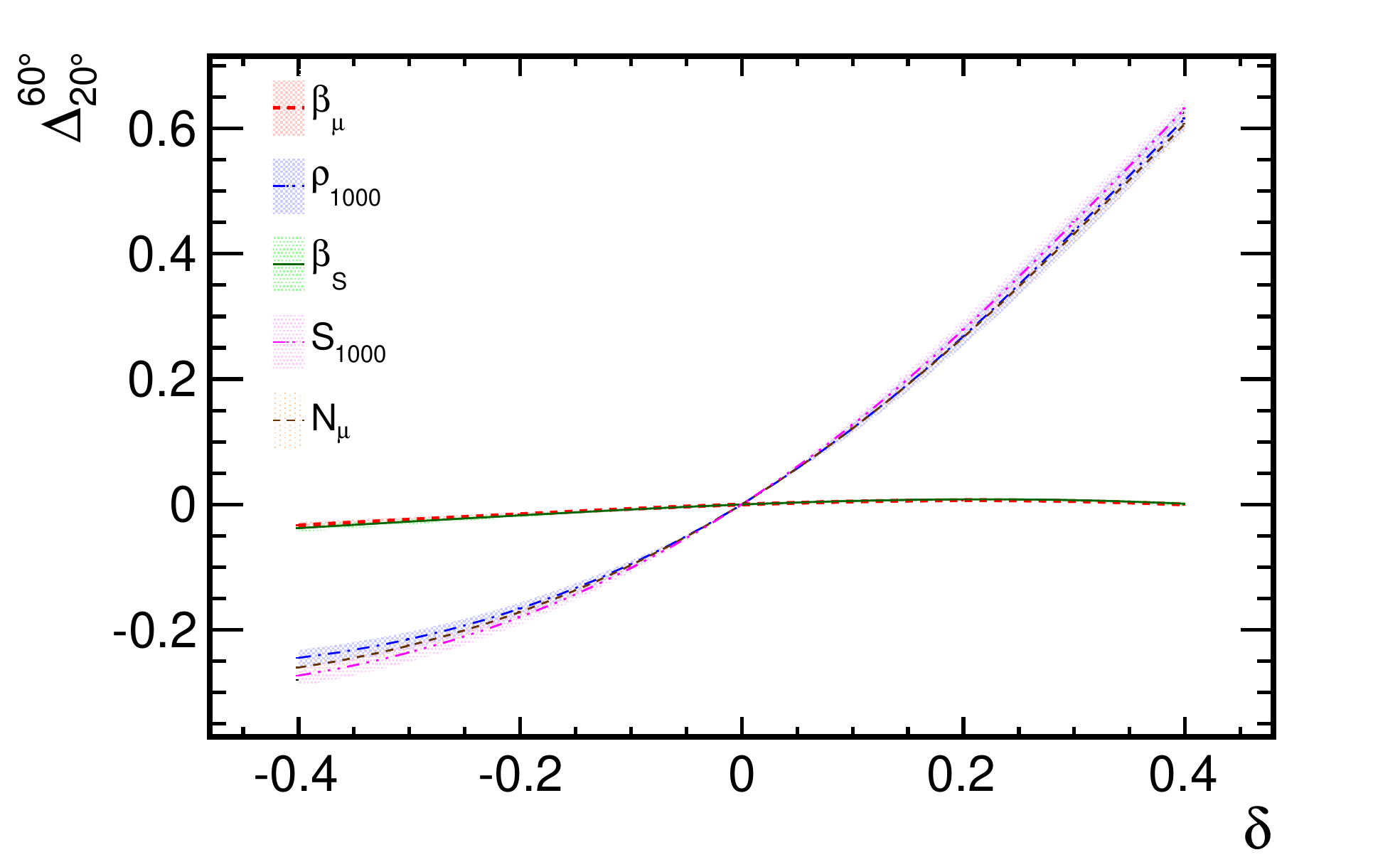}
                \caption{$\Delta_{20^\circ}^{60^\circ}\left(\delta\right)$ }
                \label{fig: DELTA1 b}
        \end{subfigure}%
\end{adjustwidth}
\caption[]{The effect of changing $\delta$ in the ratio $\frac{x\DD{}}{x\DDz}$ for the parameters described at $\theta=40^\circ$ (a) and the value $\Delta_{20^\circ}^{60^\circ}\left(\delta\right)=\frac{x_{\delta }}{x_{\delta=0}}(60^\circ) - \frac{x_{\delta }}{x_{\delta=0}}(20^\circ)$ as function of $\delta$ (b).
The shadow bands contain all compositions and models used and the line is their respective average value.
\label{fig: DELTA1}}
\end{figure}

In this section the effect of changing the muon energy spectrum with eq. \ref{eq: dNdXdE} is shown for different $\delta$ parameters. 
In figure \ref{fig: DELTA1 a}, the ratio $\frac{x\DD{}}{x\DD{=0} }$ is plotted for $x=\beta_{\mu}$, $\rho_{1000}$, $\beta_{S}$, $S_{1000}$ and $N_{\mu}$, at $\theta=40^\circ$. Both $\beta_{\mu}$ slopes change very little with the parameter $\delta$. As seen before, the size parameters change around $-35\%$ at $\delta=-0.4$ to $+20\%$ around $\delta=+0.4$. We saw that the effect on $\delta$ depends on the zenith angle, so in the figure \ref{fig: DELTA1 b} the parameter $\Delta_{20^\circ}^{60^\circ}\left(\delta\right)=\frac{x_{\delta }}{x_{\delta=0}}(60^\circ) - \frac{x_{\delta }}{x_{\delta=0}}(20^\circ)$ is plotted. For the size parameters, the modification affects more the results at $60^\circ$ than at $20^\circ$.

\subsection{Effect on the MPD}
\begin{figure}[h]
\centering
\hspace{-0.5cm}
\begin{adjustwidth}{-0.50cm}{-0.50cm}
       \begin{subfigure}[b]{0.5\linewidth}\centering
\includegraphics[width=1\textwidth]{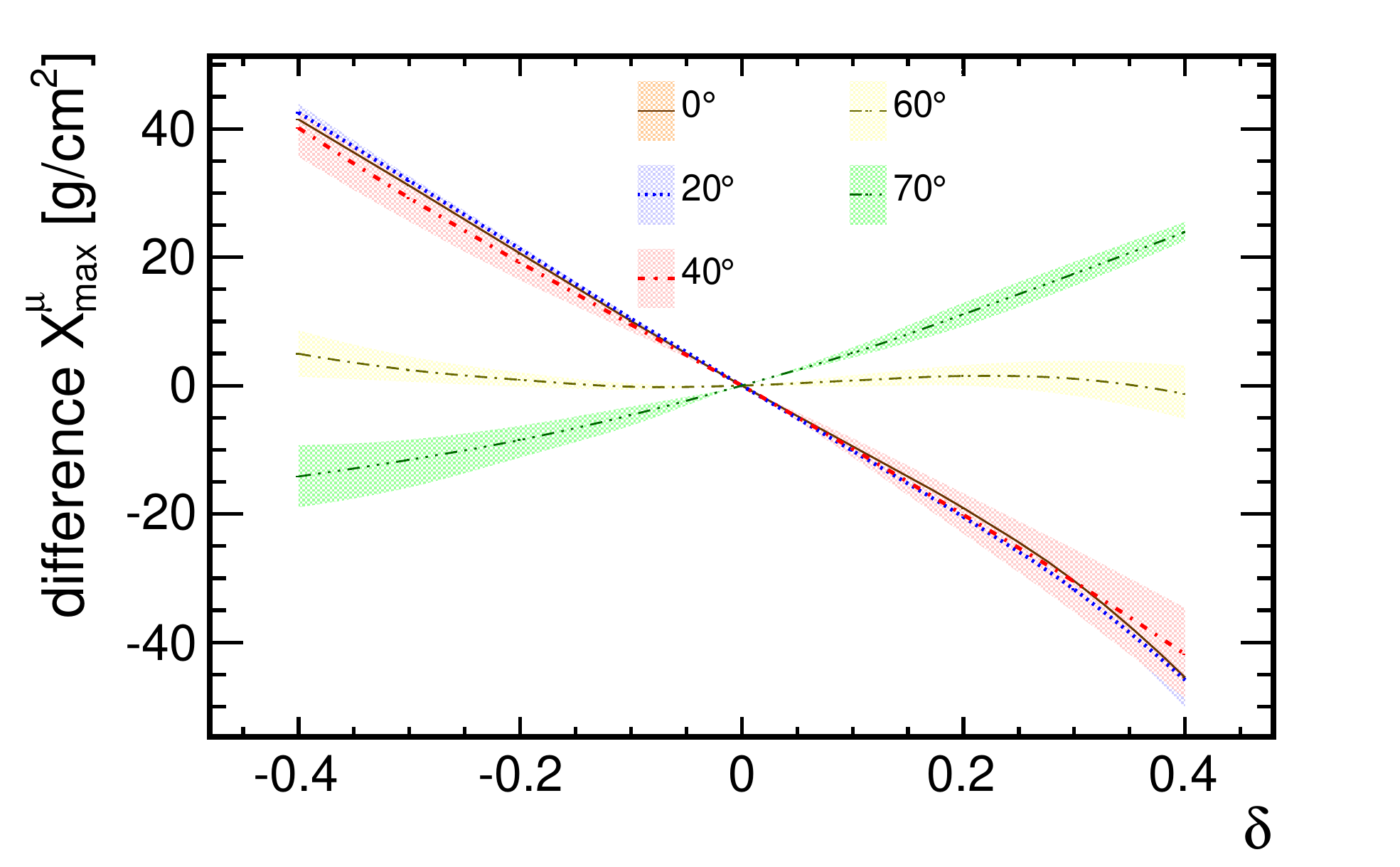}
                \caption{$\langle X_{max}^{\mu}\rangle\DD{}-\langle X_{max}^{\mu}\rangle\DDz$ }
                \label{fig: DELTA2 a}
        \end{subfigure}%
       \begin{subfigure}[b]{0.5\linewidth}\centering
\includegraphics[width=1\textwidth]{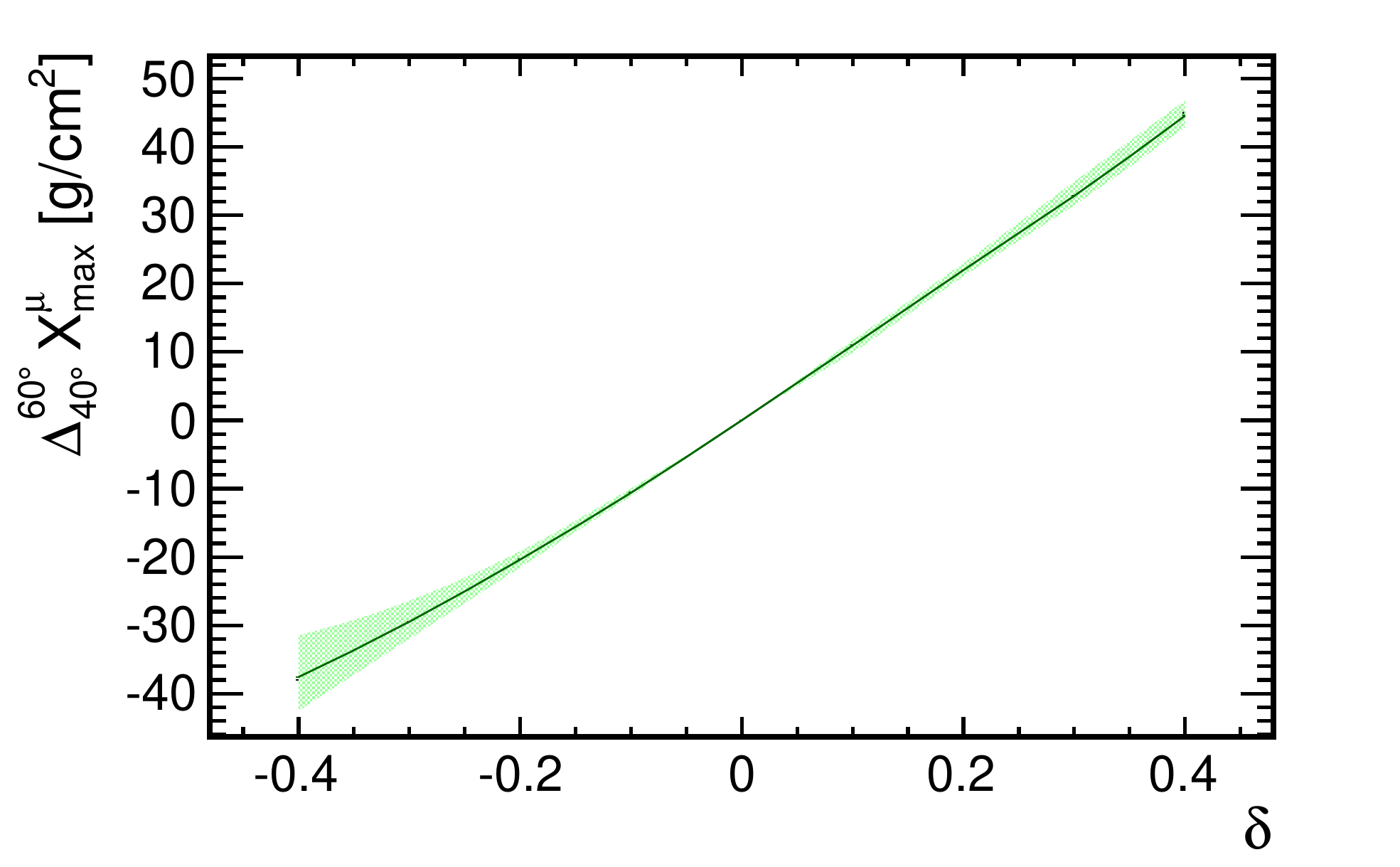}
                \caption{$\Delta_{40^\circ}^{60^\circ}\left(\delta\right) X_{max}^{\mu}$ }
                \label{fig: DELTA2 b}
        \end{subfigure}%
\end{adjustwidth}
\caption[]{The difference between $X_{max}^{\mu}$ from the modified and the unmodified spectrum, $\langle X_{max}^{\mu}\rangle\DD{}-\langle X_{max}^{\mu}\rangle\DDz$ (a) and the value $\Delta_{40^\circ}^{60^\circ}\left(\delta\right)= \langle X_{max}^{\mu}\rangle\DD{}(60^\circ) - \langle X_{max}^{\mu}\rangle\DD{}(40^\circ)$ as function of $\delta$ (b).
The shadow bands contain all compositions and models used and the line is their respective average value.
\label{fig: DELTA2}}
\end{figure}

The energy modification on the apparent $X_{max}^{\mu}$ changes drastically with zenith angle, as can be seen in figure \ref{fig: DELTA2 a}. At lower zenith angles ($\theta<60^\circ$), the apparent maximum on the MPD increases for negative $\delta$ and decreases for positive $\delta$. At $70^\circ$, we can see the contrary behaviour and at $\sim60^\circ$ the maximum is almost constant with $\delta$. This behaviour comes from different characteristic of the muon production and transport. The number of muons in the shower depends on its stage of development which is less sensitive to changes on the zenith angle for more inclined showers.
Additionally, the modification of the muon energy spectrum changes the relative importance between low energy and high energy muons (by altering the muon survival probability) producing distortions in the MPD profile that lead to subsequent differences in the $X_{max}^{\mu}$.
Finally, for more inclined showers, the propagation effects become more important and at $70^\circ$, more muons survive with positive than with negative $\delta$. 

\FloatBarrier
\section{Summary}
\label{section:Conclusions}

We have studied the impact of changing the muon energy spectrum at production on EAS observables related to the muon shower component.
The muon content at ground increases between 5\% at $20^\circ$ zenith angle to 17\% at $60^\circ$ zenith angle when the muon spectrum is changed by $E^{\delta}$, with $\delta=+0.1$.  This introduces an additional dependence on the observed number of muons at ground with respect to the zenith angle, that might be measured by experiments.

The slope of the average LDF $\beta_{\mu}$, measured through a modified NKG function for $r \in [500;2000]$ m, is relatively insensitive to variations in the energy spectrum, changing by around 2\% for $\delta=+0.1$.


The maximum of the apparent MPD $X^\mu_{max}$ changes by $10\text{ g/cm}^{2}$, and the kinematic delay changes by 3 ns independently of the distance to the core. In the MPD reconstruction algorithms, the kinematic delay is usually parametrized from models and subtracted from the total arrival time delay in order to access the pure geometric transformation from arrival times into depth. This could introduce a bias on the reconstructed apparent  $X^\mu_{max}$  of the order of $\sim 5\text{ g/cm}^{2}$ at 60 degrees and 1700
m from the core, or $10\text{ g/cm}^{2}$ at 1000 m.

The parameters studied in this work are summarised in table \ref{tab: Results}, for proton \epos simulated at $\theta=40^\circ$. Since the effect of the modified spectrum is different with zenith angle, we define the variable $\Delta_{20^\circ}^{60^\circ}(\delta)=\frac{x_{\delta}}{x_{\delta=0}}(60^\circ) - \frac{x_{\delta}}{x_{\delta=0}}(20^\circ)$, where $x_{\delta=0}$ is the parameter without modification, and it is also given in table\ref{tab: Results}.\\


It should be kept in mind, that the large number of muons on data might be caused simply by an increase in the overall number of produced muons. In that case, the shape of the muon content should be just a normalization factor at some energy for all angles. Nonetheless, if some different effect plays a role (and also due to the propagation effects in the atmosphere) the final muon normalization factor might depend on the shower zenith angle. According to \cite{GlennysICRC13}, the excess on the total signal seen in data of the Auger Observatory changes with zenith angle. This effect contains also the information about different muonic/electromagnetic ratios per angle, but it might suggest a difference in the muon normalization.  In the near future, with the upgrade of the Auger Observatory(AugerPrime)\cite{AugerPrime}, it might be possible to clarify this feature and understand where the differences in the muon component of air showers come from.

\begin{table}[h]
\centering{}
\caption[]{Summary of the variations of the parameters for proton \epos$\left(\frac{x_{\delta}}{x_{\delta=0}}\right)$, at $\theta=40^\circ$ under $\delta=\pm0.1$ modification of the spectrum, and the angular dependence of such variations as $\Delta_{20^\circ}^{60^\circ}( \delta=\pm0.1)=\frac{x_{\delta}}{x_{\delta=0}}(60^\circ)-\frac{x_{\delta}}{x_{\delta=0}}(20^\circ)$ for all variables except MPD and time, which is calculated between 40$^\circ$ and 60$^\circ$.
}
\begin{tabular}{ll|l|ll}
\toprule \toprule

         & Values & Variation ($\delta=\pm 0.1$) & Angular dependence ($\delta=\pm0.1$) & Range\\
  
 \tabularnewline  \bottomrule
$N_\mu$           & $1.57\times10^{7}$  & $\pm$8\%& $\pm$11.0\% & 20-60$^\circ$\\
$S_{1000}[\text{a.u.}]$     & 1.44   & $\pm8\%$        & $\pm$10\%   & 20-60$^\circ$\\    
$\beta_\mu$                 & 1.99   & $\pm$1.6\% & $\mp$0.7\%  & 20-60$^\circ$\\
$\beta_S$                   & 2.06   & $\pm$1.5\% & $\pm$ 0.6\% & 20-60$^\circ$\\
$X_{max}^{\mu}[\text{g/cm}^{2}]$& 611 & $\mp$10 &  $\pm$10 & 40-60$^\circ$\\
$t_\epsilon$ [ns]           & 140  &   $\mp$3   &    $\pm$0   & 40-60$^\circ$\\
 \tabularnewline
\bottomrule \bottomrule 
\end{tabular}
\label{tab: Results}
\end{table}
\normalsize

\section*{Acknowledgements}
We would like to thank J. Alvarez-Mu\~niz, C. Dobrigkeit, M. C. Esp\'irito Santo and M. Pimenta for carefully reading this manuscript. The authors wish also to thank IF/00820/2014/CP1248/CT0001, SFRH/BPD/73270/2010, SFRH/BD/89109/2012, OE, FCT-Portugal and CERN/FIS-NUC/0038/2015 for financial support. 


\section*{References}
\bibliography{myref}

\FloatBarrier
\appendix

\end{document}